\documentclass{aa}  
\usepackage{multirow}
\usepackage{stackengine}
\newcommand\xrowht[2][0]{\addstackgap[.5\dimexpr#2\relax]{\vphantom{#1}}}

\usepackage{graphicx}
\usepackage{txfonts}
\usepackage[dvipsnames]{xcolor}
\usepackage{orcidlink}
\usepackage{colortbl}
\usepackage[rightcaption]{sidecap}
\sidecaptionvpos{figure}{c}

\usepackage{hyperref}  
\hypersetup{colorlinks=true,linkcolor=[rgb]{1.,0.2,0.2},citecolor=[rgb]{0.1,0.4,1.},filecolor=[rgb]{0.7,0.2,0.2},urlcolor=[rgb]{0.7,0.2,0.2}}

\usepackage{color}
\definecolor{blue}{rgb}{0., 0., 1}

\def\cgs{\rm{erg}\ \rm{s^{-1}}\ \rm{cm^{-2}}}
\def\ergs{\rm{erg}\ \rm{s^{-1}}}
\def\Msunyr{\rm{M}_\odot\ \rm{yr}^{-1}}
\def\Lsun{L$_\odot$}
\def\Msun{M$_\odot$}

\newcommand{\NeIV}{{[Ne\,{\sc iv}]\,}}

\newcommand{\NeV}{{[Ne\,{\sc v}]\,}}

\newcommand{\OIII}{{[O\,{\sc iii}]\,}}

\newcommand{\OIIIl}{{[O\,{\sc iii}]\,$\lambda$}}

\newcommand{\CIIfir}{{[C\,{\sc ii}]$\rm \lambda158\mu m\,$}}
\newcommand{\CII}{\text{[C\,{\sc ii}]\,}}

\newcommand{\CIV}{{C\,{\sc iv}\,}}

\newcommand{\HeII}{{He\,{\sc ii}\,}}

\newcommand{\Ha}{H$\alpha$\,}

\newcommand{\Hb}{H$\beta$\,}

\def\kms{km\,s$^{-1}$}

\def\jykms{Jy~km\,s$^{-1}$}

\begin{document} 

\authorrunning{G. Mazzolari et al.}
\titlerunning{NOEMA non-detection of JWST-AGN at $z>6$}
\title{Constraints on the host galaxy and AGN properties of three $z > 6$ JWST AGN from NOEMA observations}
\author{
Giovanni Mazzolari\inst{1}\thanks{ \email{gmazzolari@mpe.mpg.de}}$^{\orcidlink{0009-0005-7383-6655}}$, 
Hannah Übler\inst{1}$^{\orcidlink{0000-0003-4891-0794}}$, 
Rodrigo Herrera Camus\inst{2,3}$^{\orcidlink{0000-0002-2775-0595}}$,
Ric Davies\inst{1}$^{\orcidlink{0000-0003-4949-7217}}$,
Linda Tacconi\inst{1}$^{\orcidlink{0000-0002-1485-9401}}$,
Dieter Lutz\inst{1}$^{\orcidlink{0000-0003-0291-9582}}$,
Natascha F\"orster Schreiber\inst{1}$^{\orcidlink{0000-0003-4264-3381}}$,
Francesco D'Eugenio\inst{4,5}$^{\orcidlink{0000-0003-2388-8172}}$,
Minju Lee \inst{6,7}$^{\orcidlink{0000-0002-2419-3068}}$,
Capucine Barfety \inst{1}$^{\orcidlink{0000-0002-1952-3966}}$,
Elena Bertola \inst{8}$^{\orcidlink{0000-0001-5487-2830}}$,
Andrew Bunker \inst{9}$^{\orcidlink{0000-0002-8651-9879}}$,
Andreas Burkert \inst{10,1}$^{\orcidlink{0000-0001-6879-9822}}$,
Jianhang Chen \inst{1}$^{\orcidlink{0000-0003-3921-3313}}$,
Giovanni Cresci \inst{8}$^{\orcidlink{0000-0002-5281-1417}}$,
Frank Eisenhauer \inst{1,11},
Juan Manuel Espejo Salcedo  \inst{1}$^{\orcidlink{0000-0001-6703-4676}}$,
Simon Flesch  \inst{1}$^{\orcidlink{0009-0003-5925-0354}}$,
Reinhard Genzel \inst{1,10,13}$^{\orcidlink{0000-0002-2767-9653}}$,
Xihan Ji \inst{4,5}$^{\orcidlink{0000-0002-1660-9502}}$,
Lilian Lee \inst{1},
Daizhong Liu \inst{1,15}$^{\orcidlink{0000-0001-9773-7479}}$,
Cosimo Marconcini \inst{8}$^{\orcidlink{0000-0002-3194-5416}}$,
Roberto Maiolino \inst{4,5,12}$^{\orcidlink{0000-0002-4985-3819}}$,
Thorsten Naab \inst{16}$^{\orcidlink{0000-0002-2864-9560}}$,
Amit Nestor Shachar \inst{17}$^{\orcidlink{0000-0003-1785-1357}}$,
Meghana Pannikkote \inst{1}$^{\orcidlink{0009-0004-8724-9163}}$,
Eleonora Parlanti \inst{14}$^{\orcidlink{0000-0002-7392-7814}}$,
Stavros Pastras \inst{1}$^{\orcidlink{0009-0009-0472-6080}}$,
Michele Perna \inst{18}$^{\orcidlink{0000-0002-0362-5941}}$,
Claudia Pulsoni \inst{1}$^{\orcidlink{0000-0002-1428-1558}}$,
Bruno Rodríguez del Pino \inst{18}$^{\orcidlink{0000-0001-5171-3930}}$,
Eckhard Sturm \inst{1}$^{\orcidlink{0000-0002-0018-3666}}$,
Taro Shimizu \inst{1}$^{\orcidlink{0000-0002-2125-4670}}$,
Giulia Tozzi \inst{1}$^{\orcidlink{0000-0003-4226-7777}}$
}
\institute{
Max-Planck-Institut für extraterrestrische Physik, Gießenbachstraße 1, 85748 Garching, Germany.
\and
Departamento de Astronomía, Universidad de Concepción, Concepción, Chile.
\and 
Millennium Nucleus for Galaxies, Concepción, Chile.
\and 
Kavli Institute for Cosmology, University of Cambridge, Madingley Road, Cambridge CB3 0HA, UK. 
\and 
Cavendish Laboratory – Astrophysics Group, University of Cambridge, 19 JJ Thomson Avenue, Cambridge CB3 0HE, UK 
\and
Cosmic Dawn Center (DAWN), Denmark
\and
2DTU-Space, Technical University of Denmark, Elektrovej 327, DK2800 Kgs. Lyngby, Denmark
\and
INAF–OAA, Osservatorio Astrofisico di Arcetri, largo E. Fermi 5, 50127, Firenze, Italy
\and
Department of Physics, University of Oxford, Denys Wilkinson Building, Keble Road, Oxford OX13RH, U.K.
\and 
University Observatory, Ludwig Maximilians University, Scheinerstr. 1, 81679 Munich, Germany.
\and 
Department of Physics, TUM School of Natural Sciences, Technical University of Munich, 85748 Garching, Germany.
\and 
Department of Physics and Astronomy, University College London, Gower Street, London WC1E 6BT, UK
\and 
Departments of Physics \& Astronomy, University of California, Le Conte Hall, Berkeley, CA, 94720, USA.
\and
Scuola Normale Superiore, Piazza dei Cavalieri 7, I-56126 Pisa, Italy
\and
Purple Mountain Observatory, Chinese Academy of Sciences, 10 Yuanhua Road, Nanjing 210023, China
\and
Max-Planck-Institut für Astrophysik, Karl-Schwarzschild Straße 1, 85748 Garching, Germany
\and
School of Physics and Astronomy, Tel Aviv University, Tel Aviv 69978, Israel
\and
Centro de Astrobiología (CAB), CSIC-INTA, Ctra. de Ajalvir km 4, Torrejón de Ardoz, E-28850, Madrid, Spain
}
   \date{}


    \abstract
  { The \textit{James Webb Space Telescope} revealed a large population of active galactic nuclei (AGN) at $z>4$, showing peculiar physical properties that are challenging to reconcile with known lower redshift and higher luminosity AGN. A missing piece in the description of these sources is the physical characterisation of their host galaxies, and a reliable modeling of the emission coming from the AGN and from the host. 
  We targeted with deep NOEMA observations the \CIIfir emission of three {\it JWST}-discovered AGN at $z>6$. 
  Two of them have the typical features of Little Red Dots (LRDs), while the third one is a blue, extended, Type I AGN. We do not significantly detect \CII\ emission or dust continuum in any of the targets, even after stacking. The resulting \CII\ luminosity upper limits, $\log (L_{[CII]}/$\Lsun$)<7.77-8.1$, lie $\sim2\sigma$ below the values expected from the \CII–SFR relation, and we explore different scenarios to explain the lack of \CII. We obtained upper limits on the gas masses of $\log (M_{gas}/$\Msun$)<9.26-9.59$ corresponding to $\log( M_{dust}/$\Msun$)<5.68-6.55$ assuming a metallicity dependent dust to gas ratio. Using the continuum non-detections ($\rm rms\sim 16-25 ~\mu Jy$) together with {\it JWST}/MIRI constraints, we performed a revised SED-fitting decomposition, resulting in stellar masses up to $\sim 2$ dex lower than previously reported, and implying $0.03\lesssim M_{BH}/M_{*}\lesssim0.7$. For the two LRDs, the SED is well reproduced by stellar emission in the rest-frame UV, while the rising rest-frame optical slope, flattening toward the near-infrared, is consistent with emission from a Type I AGN partially obscured along the polar direction with $E(B-V)_{\rm polar}\simeq1$, in agreement with attenuation derived from the broad lines Balmer decrement. This decomposition demonstrates that a relatively standard AGN configuration can reproduce the SEDs of the two LRDs, without invoking more exotic scenarios. Finally, we investigate the positions of the three sources in the $IRX-\beta_{UV}$ plane, finding that they lie in a parameter space where galaxies are typically characterized by patchy dust distributions. Our analysis highlights the importance of millimeter constraints to characterize the different physical properties of high-z AGN.}





   \keywords{ galaxies: active, galaxies: high-redshift,  galaxies: ISM,  galaxies: nuclei, submillimeter: galaxies 
               }

   \maketitle
%

\section{Introduction}\label{sec:intro}

One of the most exciting results of the first years of {\it JWST} operations is the discovery of a surprising abundance of broad-line (BL) active galactic nuclei (AGN) (also called Type I AGN) at $z>4$, characterized by bolometric luminosities $43<\log (L_{bol}/ \rm erg\ s^{-1})<46$ that are 1-3 dex fainter than the previously known QSO population \citep{Matthee23, Maiolino2024_AGNsample, Harikane23, Greene23, Juodzbalis25_AGNsample, Taylor2024_AGN, Hviding25}. 
This new population of faint Type I AGN is characterized by peculiar physical properties, rarely observed in the previously known AGN population. Specifically, they are characterized by a weakness in X-ray \citep[with observed $L_{2-10 keV}$ up to 2-3 dex lower than expected, see][]{Maiolino24_X, Yue24, Ananna24_Xweak, Mazzolari25_CEERS, Comastri25} and radio emission \citep{Mazzolari26_radioJWST,Gloudemans25} compared to what is expected from the usual scaling relations derived from lower redshift and higher luminosities AGN \citep{duras20,Lusso12,damato22,Wang24_FP,Bariuan22}. Additionally, the black holes (BHs) powering these sources appear to be over-massive compared to their host galaxies, when placed on local scaling relations \citep{Ubler23,Kokorev23,Furtak23, Maiolino2024_AGNsample,Tripodi25_massiveBH,Napolitano25_BH, Pacucci23,Jones25_MBHMstar}, such as those by \cite{Reines15, Kormendy13}.  While there are still uncertainties on both the BH and stellar mass measurements for these objects, this now appears to be a systematic trend, and some of these sources even reach $M_{BH}/M_{*}>1$ \citep{Juodzbalis25_Mdyn}. 
Additionally, $\sim 20\%-30\%$ of these high-z AGN discovered by {\it JWST} are characterized by narrow absorptions in the broad hydrogen and helium emission lines \citep{Matthee23,Juodzbalis2024b,Wang25_LRDz3,Juodzbalis25_AGNsample,Deugenio26_restabs, Loiacono25_bird}, a spectroscopic feature characterizing only $\sim0.1\%$ of local AGN \citep{Hutchings02_narrowabs, Wang15_narrowabs, Burke21_narrowabs,Shi16_narrowabs, Schultze18_narrowabs}.

Among these {\it JWST}-discovered Type I AGN, a subclass, consisting of $\sim 30-50\%$ of the population \citep[the fraction significantly depend on the luminosity of these objects, see ]{Hviding25, Hainline2024}, are the so-called Little Red Dots \citep[LRDs][]{Matthee23}. These sources are characterized by all the features reported above but specifically also by very compact morphologies, typically unresolved at the limit of {\it JWST} PSF (i.e. sizes $\lesssim 0.1^{\prime\prime}$) and by a spectral energy distribution (SED) with a peculiar V-shaped profile (i.e., a red slope in the rest-frame optical and a blue slope in the rest-frame UV) presenting a turnover in correspondence of the Balmer break \citep{Kocevski2024_LRD, Greene23, Setton25_Vshape, Ji25_QSO1a}.
Different works have tried to explain the unexpected physiscal properties of {\it JWST} discovered AGN and LRDs, but many pieces are still missing. Several studies are now suggesting the existence of a dense cocoon of gas surrounding the central active BH \citep{Inayoshi2024_densegas,Ji25_QSO1a, Naidu25_BHstar, deGraaff25_cliff,Inayoshi25_LRDreview}, a scenario first proposed by \cite{Inayoshi2024_densegas} to explain the origin of the V-shaped SED. If the rising SED redwards of the Balmer break arose from stars, then these sources would have stellar densities higher than those of the most dense stellar clusters \citep{Baggen24_starcluster}, and their number density would challenge the $\Lambda$CDM model \citep{Akins25_LRD}. The same ``V-shape" SED, however, could result from hydrogen absorption by a dense gas envelope surrounding the central BH. If the gas density is sufficiently high to keep the hydrogen $n=2$ level populated via collisional excitation, then the Balmer break emerges as a direct consequence of this AGN structure (through absorption in all Balmer lines), while the X-ray and radio weakness would instead be attributed to absorption by the large column densities surrounding the central emitting source (i.e. from photoelectric absorption and free-free absorption, respectively). This scenario has also been supported by the identification of LRDs with Balmer break strengths exceeding the expected limit for stellar populations \citep{Naidu25_BHstar, deGraaff25_cliff}. More recently, some works also proposed the so-called "BH-star" model, where the dense gas envelope surrounding the central BH effectively acts as the convective envelope of a star, and where scattering processing can also be in place in an inner ionizied layer, producing exponential wings in the broad emission lines and falsifying the measure of BH masses based on local virial relations \citep{Naidu25_BHstar, Torralba25}. In this scenario, the dense gas envelope would behave as the thermalized photosphere of a star, emitting a blackbody radiation at temperatures $T\sim5000$K, peaking at $\lambda_{rest}\sim0.65 \mu m$, and producing the optical rising continuum \citep{Lin26_egg,Kido25_BHstar,deGraaff25_BB, Barro25}.

A missing piece in the physical characterisation of {\it JWST}-discovered AGN and LRDs are the properties of their host galaxies.
Their small sizes make it difficult to disentangle the host galaxy and AGN contributions in the optical bands, and a natural way to unambiguously observe the host galaxy would be to examine its millimeter emission. However, most previous attempts to detect the \CIIfir or dust continuum in {\it JWST}-discovered AGN at z$>$4 led to non-detections \citep{Labbe25,Akins25_LRD,Setton25_dust,Xiao25_noema}. While a tentative [CI](2–1) detection was reported for one LRD \citep{Akins25_CI}, the only \CIIfir line detection in a {\it JWST}-discovered AGN reported so far is for A383-LRD1 \citep{Golubchik25}, which also shows \OIIIl88$\mu$m emission \citep{Knudsen25}. However, the LRD nature of this source is just suggested by its NIRCam color, while the presence of BL emission and possibly other LRD spectroscopic signatures are not yet confirmed. Additionally, it also shows a close-by blue companion, and both the \CII and \OIIIl88$\mu$m seem to peak on the companion or in between the two sources, suggesting that these lines might actually trace a bridge of common circumgalactic medium rather than the LRD host. This prevents a measure of the dynamical mass of this LRD, but the high \OIII/\CII ratio suggests a strong UV radiation field coming from the sources.

FIR observations would also be crucial to constrain the dust mass budgets of {\it JWST}-discovered AGN. While quasars at $z\sim 7$ do show similar dust properties to their lower redshift analogs \citep[both from a millimeter and mid-infrared perspective,][]{Salvestrini25_mmQSO, Bosman25}, no dust continuum detection has been reported so far in the mm band for high-z AGN discovered by {\it JWST}, and stacking experiments put a median upper limit to the dust masses of their hosts of $M_{dust}\lesssim10^6$\Msun~ at $z\sim6$ \citep{Labbe25_LRDsample,Casey25_Mdust, Casey24_LRDdust}. At the same time, the hot dust contribution from the dusty torus is found to be minor, if not absent, in some single sources \citep{Setton25_dust,Wang25_LRDz3, Williams24_lacktorus}, but detected in others \citep{Brazzini26_LBDLRD} and detected also in some stacking experiments \citep{Delvecchio25}.  In general, these indications seem to favor the scenario in which both the warm and the cold dust content of these sources are small. Some works further proposed that the Balmer decrements observed in some of these objects (reaching up to \Ha/\Hb$\sim10$) are not due to dust obscuration effects but rather to different physical conditions (deviating from Case B recombination), or due to scattering effects \citep{Chang25_scattering, deGraaff25_BB}.

Recently \cite{Greene25} showed that the SED of two of the brightest LRDs is systematically different from the standard, nearby Type I AGN SED. While \cite{Greene25} focused on only two LRDs, it still suggested the need for new bolometric corrections to derive their bolometric luminosity. Crucially, the current lack of detections in the FIR of these sources makes these new bolometric corrections quite uncertain, varying by a factor of $\sim 10$ depending on how the current FIR upper limits are taken into account in the LRD SED modelling. Therefore, obtaining FIR constraints (detections or deep upper limits) on the FIR emission of these sources is important for understanding their physical properties and correctly interpreting their nature \citep[see also:][]{Herrera-Camus26_JWSTALMA}. 

In this work, we report and discuss the \CII and dust continuum upper limits of three {\it JWST}-discovered AGN, as inferred from deep NOEMA observations, and we investigate the implications of these non-detections on the AGN and host galaxy physical properties.
The paper is organized as follows. In Section~\ref{sec:target}, we present and analyze the targets and the NOEMA observations. In Section~\ref{sec:results} we present the main results from the analysis of the NOEMA non-detections. In Section~\ref{sec:undetections} we derive the \CII and continuum upper limits, in Section~\ref{sec:CIIund} we present the possible scenarios leading to a \CII non-detection, and we derive gas and dust mass upper limits in Section~\ref {sec:Mgas_Mdust_ul}. In Section~\ref{sec:dustund} we perform a new SED-fitting decomposition of the three sources using \texttt{CIGALE}, taking into account the NOEMA continuum upper limits and also JWST/MIRI data, finding important constraints on their AGN emission and properties (Section~\ref{sec:AGN_fit}). In Section~\ref{sec:IRX}, we place the three targets on the IRX-$\beta$ diagram to gain insights into their geometries and dust properties. Finally, in Section~\ref{sec:conclusion} we summarize the results and conclude.

   

\section{Targets analysis and NOEMA observations}\label{sec:target}

\begin{figure*}[h!]
    \centering
    \includegraphics[width=0.85\linewidth]{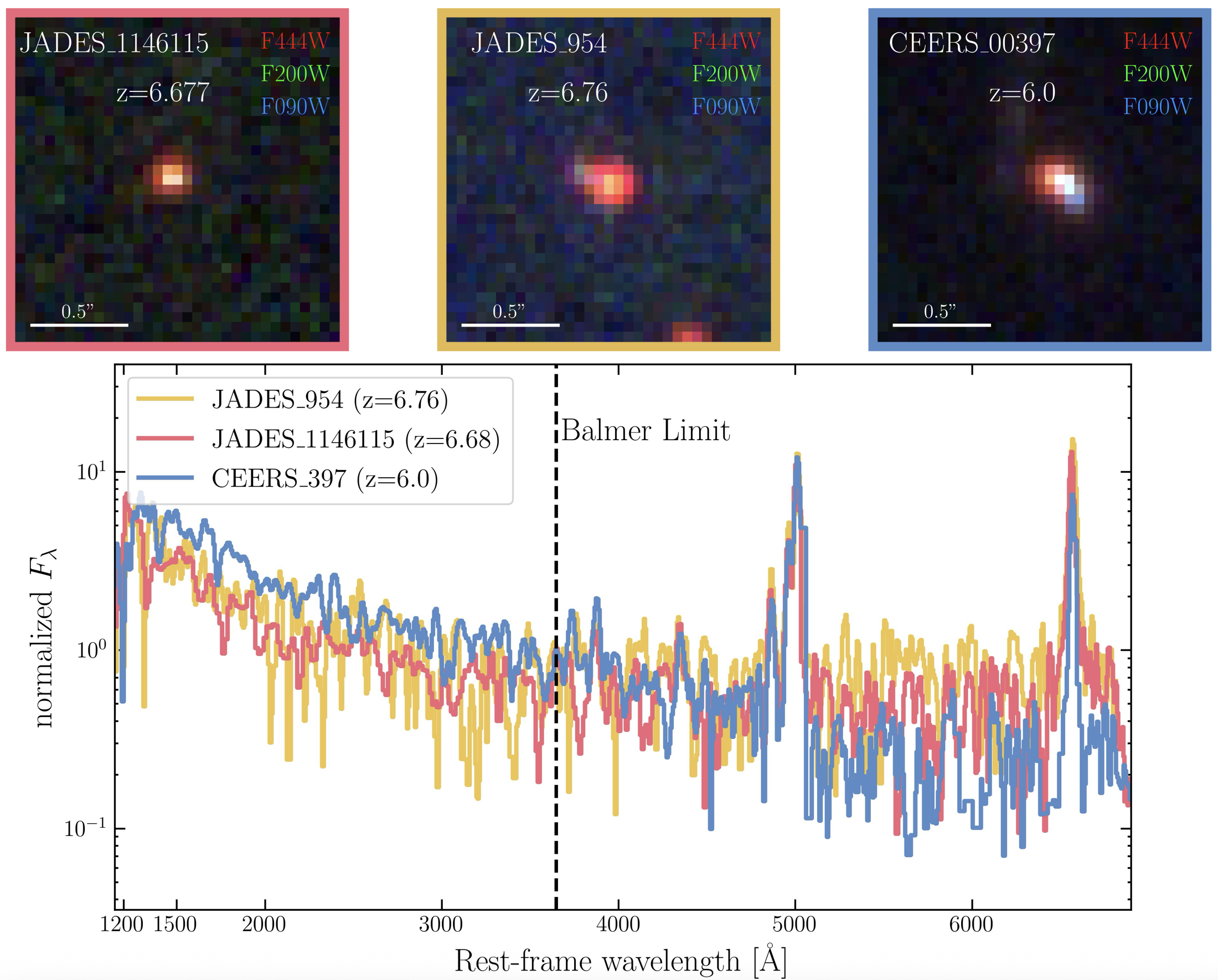}
    \caption{{\it Top:} {\it JWST}/NIRCam RGB cutouts of our targets obtained from publicly available imaging from the {\it JWST}/DAWN archive. {\it Bottom:} NIRSpec/PRISM spectra obtained from the BlackTHUNDER program (for JADES\_954 and CEERS\_00397) and from JADES DR3 for JADES\_1146115. The spectra are normalized at the Balmer limit to emphasize the different $\beta_{opt}$ at redder wavelengths. JADES\_1146115 is a LRD, JADES\_954 has the V-shape but is slightly resolved, while CEERS\_00397 is an extended blue Type I AGN.}
    \label{fig:targets}
\end{figure*}

\begin{table}[h!]
\caption{Physical properties of the three targets derived from the literature (when specified) or from the analysis outlined in Sect.~\ref{sec:targets_JWST}.}
\centering
\setlength{\tabcolsep}{2pt}
\begin{tabular}{l c c c}

Parameter & JADES\_954 & JADES\_1146115 & CEERS\_397\\ 

\hline
\\

RA [deg]& 189.151958 & 189.091458 & 214.836208 \vspace{0.15cm}\\

DEC [deg] & 62.2596 & 62.228111 & 52.882694 \vspace{0.15cm}\\

$z$ & 6.767 & 6.68 & 6.00 \vspace{0.15cm}\\

$\rm FWHM_{NL}$ [\kms] & 128 & 180 & 240 \vspace{0.15cm}\\

$\rm \log (M_{BH} /{M_\odot})$ & 7.74$^{+0.31}_{-0.32}$ & 8.57$^{+0.37}_{-0.38}$ & 7.29$^{+0.14}_{-0.14}$ \\
literature &  &  &  \vspace{0.15cm}\\

$\rm \log (M_{*} /{M_\odot})$ & 10.7 & 8.9 & 9.4 \\
literature &  &  &  \vspace{0.15cm}\\

$M_{BH}/M_*$ & 0.001 & 0.46 & 0.007 \\
literature &  &  &  \vspace{0.15cm}\\

$\rm SFR\ [\Msunyr]$ & 25 & 45 & 73 \\
from \Ha &  &  &  \vspace{0.15cm}\\

$E(B-V)$ & $\sim$0 & 0.62 & 0.12 \vspace{0.15cm}\\

$Z$ $[Z_{\odot}]$& 0.12 & 0.04 & 0.15 \\
literature &  &  &  \vspace{0.15cm}\\

$L_{AGN, bol}$ $[\ergs]$ & $1.2^{+0.6}_{-0.5}\times10^{45}$ & $2^{+7}_{-3}\times10^{44}$ & $3.3^{+1}_{-0.45}\times10^{44}$ \\
literature &  &  &  \vspace{0.15cm}\\

$R_{eff}$ $[pc]$& 357 & <270 & 658 \vspace{0.15cm}\\

Reference & 1 & 2 & 3 \\

\hline
\end{tabular}

\tablefoot{The table reports the main physical properties of the three sources as derived from the literature or from the new {\it JWST} data. The SFR is derived from the narrow \Ha\ flux by correcting for dust attenuation and using the scaling relation reported in \cite{Kennicutt12}. The $E(B-V)$ values for JADES\_954 and CEERS\_397 have been updated relative to those reported in the literature, leveraging new {\it JWST} high-resolution spectra from the BlackTHUNDER program. The sources' effective radii are measured using \texttt{pysersic} as outlined in Sect.~\ref{sec:targets_JWST}. The literature references are as follows: (1) \cite{Maiolino2024_AGNsample}, (2) \cite{Juodzbalis24}, (3) \cite{Harikane23}.}

\label{tab:sample}
\end{table}

\subsection{{\it JWST} data and literature}\label{sec:targets_JWST}
Our sample consists of three Type I AGN: JADES\_954 \citep{Maiolino2024_AGNsample}, JADES\_1146115 \citep{Juodzbalis24}, and CEERS\_397 \citep{Harikane23}. These $6<z<7$ targets were chosen from spectroscopically confirmed Type I AGN discovered by {\it JWST},and observable with NOEMA in the Northern hemisphere. 
The broad emission component in the Balmer lines (specifically in the \Ha) attributed to BLR emission of a Type I AGN was detected in the {\it JWST}/NIRSpec medium-resolution (MR, $R\sim1000$) spectra for JADES\_954 and CEERS\_397, and in the {\it JWST}/NIRSpec PRISM spectrum for JADES\_1146115.
For all sources, we uniformly recomputed the BH masses and AGN bolometric luminosities using the locally calibrated relations from \cite{Reines15} and \cite{stern12}, respectively, given the lack of corresponding $z\sim6$ scaling relations. 
In addition to the available {\it JWST}/NIRSpec MR spectra, JADES\_954 and CEERS\_397 were observed as part of the NIRSpec-IFS Large Program BlackTHUNDER (PID: 5015; PIs: H.~\"Ubler, R.~Maiolino), providing prism ($R\sim100$) and high-resolution (HR, $R\sim2700$) integral-field unit (IFU) spectroscopy. 
The global physical properties of the three targets, as derived from previous literature works and from the additional analysis presented in this section, are summarized in Table~\ref{tab:sample}. In Fig.~\ref{fig:targets} we show their {\it JWST} NIRSpec/PRISM spectra and NIRCam RGB cutouts.

Our targets were selected without requiring them to be LRDs, but we characterise them following the LRD criteria outlined by  \cite{Hviding25}: broad Balmer lines, a V-shaped SED, and a compact morphology. \cite{Hviding25} showed that once two of these criteria are present, $>80\%$ of them also fulfill the third. All our targets already fulfill the first condition, as they all show broad \Ha emission.
We fitted the available {\it JWST}/PRISM spectra with a broken power-law model (with the turnover in correspondence to the Balmer limit) to compute the rest-frame optical and UV slopes. Then, we also checked the compactness criterion in the reference F444W NIRCam band \citep[where the AGN emission is expected to dominate at these redshifts, see][]{Hviding25, deGraaff25_BB}. We first measured the radial surface brightness profiles using the \texttt{photutils.profiles} module \citep{Bradley25_photutils} with circular apertures that are half pixel wide, ranging from 0 to 15 pixels, and we compared it with the profile obtained from \texttt{stpsf} \cite{Perrin14_stpsf} for the same filter (with \texttt{oversampling=4}) and applying the \texttt{jitter\_sigma} corrections reported in \cite{Morishita24_pysersic}. Second, we utilized \texttt{pysersic} \citep{Pasha23_pysersic} to model their emission in the F444W band as the sum of a central PSF-like component plus an extended stellar emission described by a Sersic profile. The results of the LRD analysis and the physical properties of the three sources are reported below.

\subsubsection{An extended LRD: JADES\_954}
Our first target, JADES\_954 at $z=6.76$ is the most luminous Type I AGN in the selection performed by \cite{Maiolino2024_AGNsample}. Its host galaxy stellar mass was estimated to be $\log(M_{*}/M_\odot)=10.7$ based on a spectral fit performed with \texttt{BEAGLE-AGN} \citep{VidalGarcia24} decomposing the emission coming from the stellar population and the one coming from a reddened AGN. From the literature results, this source has a $M_{BH}/M_{*}=0.001$, and it is only slightly above the $M_{BH}-M_{*}$ relation as derived from locally calibrated, single-epoch scaling relations \citep{Reines15,Volonteri16_MBHMstar}. \cite{Maiolino2024_AGNsample} also computed a narrow line $E(B-V)=0.23$ using {\it JWST} MR spectra obtained with NIRSpec-MSA. Using the BlackTHUNDER HR spectrum extracted from NIRSpec-IFU observations, we found \Ha/\Hb$\sim 2.83$, consistent with the nominal value of 2.86 for case B recombination, implying no attenuation in the narrow line in the rest-UV and optical. 
This can likely be attributed to the different spectral decomposition based on the MR and HR data.
Both broad \Ha\ and \Hb\ are detected. From the broad Balmer decrement we found \Ha$_{broad}$/\Hb$_{broad}\sim9$, corresponding to to $E(B-V)=1.1$ (or to $A_V\sim3$) assuming a Small Magellanic Cloud extinction law \citep[SMC,][]{Gordon03}. 

Additionally, the high-S/N detection of the broad \Ha emission and the high spectral resolution revealed a clear narrow absorption feature $\sim200$\kms\ blueshifted compared to the systemic redshift (Mazzolari et al.~in prep.). This feature is observed in at least $\sim 10\%-20\%$ of the {\it JWST}-discovered AGN \citep{Kocevski2024_LRD, Matthee23, Lin26_egg} and was not detected in the NIRSpec MR spectrum, supporting that HR spectroscopy is needed to reveal these narrow absorptions \citep{Deugenio26_restabs}.

From the UV and optical spectral slope fitting we obtained for JADES\_954 $\beta_{UV}=-2.10$ and $\beta_{opt}=0.72$, fulfilling the `V-shape' criterion for LRDs. Regarding compactness, the radial surface brightness profile of JADES\_954 was slightly extended compared to the PSF profile in the F444W filter, while the \texttt{pysersic} fit returned a non-negligible PSF-like component, but not dominant compared to the total source flux ($f_{PSF}=0.37$). This result is mostly due to the presence of a small blue extension of the source in the North-Est direction, as often observed also for other LRDs \citep{Rinaldi25_LRD}. From the \texttt{pysersic} fit, we derived an effective radius of $\sim 350$pc. Given that two out of the three LRD criteria are satisfied (especially the V-shape SED), and the extension compared to the PSF is only marginal, we considered this source as a marginally extended LRD. Searching for counterparts in the available {\it JWST}/MIRI F1000W and F2100W images of GOODS-N (Program ID: 5407, PI: G. Leung), we found a detection in the F1000W image but not in F2100W. To define the MIRI $3\sigma$ upper limit, we followed the same approach as in \cite{Alberts24_smiles}, multiplying the local image rms by the empirically derived F2100W PSF area and then multiplying by 1.5 to account for correlated noise.

\subsubsection{LRD: JADES\_1146115}
Our second target, JADES\_1146115 is the `dormant' BH  at $z=6.67$ presented and discussed by \cite{Juodzbalis24} \citep[also referred to as GN-1001830 or JADES-GN-38509 in][]{Juodzbalis25_AGNsample} and observed with NIRSpec PRISM and medium resolution spectroscopy as part of the JADES survey \citep{Bunker20_JADESspecsurvey,Eisenstein23}. The stellar mass of $\log(M_{*}/M_\odot)=8.9$ was derived by performing first a 2D image decomposition that separated the unresolved nuclear source (AGN) from the compact host galaxy with a disk-like profile and then by fitting just the host galaxy emission with both \texttt{BAGPIPES} \citep{Carnall19_bagpipes} and \texttt{Prospector} \citep{Johnson21_prospector}. This source is clearly offset from the local $M_{BH}/M_{*}$ scaling relations, hosting an overmassive BH with $M_{BH}/M_{*}=0.47$. The narrow line attenuation taken from \cite{Juodzbalis24} is $E(B-V)=0.62$. While the broad \Ha emission is clearly detected in the JADES PRISM spectrum,  broad \Hb is not detected in the JADES medium resolution spectrum.

The continuum of this source fulfils the V-shape criterion, having $\beta_{UV}=-2.15$ and $\beta_{opt}=0.34$. Additionally, it has an F444W brightness profile consistent with the PSF; it is therefore unresolved and fulfills all LRD criteria. Instead, at shorter wavelengths an extended component starts to dominate the emission \citep{Juodzbalis24}. From the F444W  \texttt{pysersic} 2D fit in the F444W image we derived an upper limit to the effective radius $<$270 pc. This source is not detected in any of the MIRI F1000W and F2100W images covering the GOODS-N field (as expected, being $\sim1$ dex fainter than JADES\_954).

\subsubsection{A blue, Type I AGN: CEERS\_397}
Finally, our third target CEERS\_397 at $z=6.00$ was selected as a Type I AGN from the CEERS \citep{Finkelstein25_CEERS} {\it JWST}/MR spectrum by \cite{Harikane23}. They derived a host galaxy stellar mass of $\log(M_{*}/M_\odot)=9.4$ by first performing a 2D decomposition by fitting the image with the AGN and host galaxy component and then using \texttt{Prospector} to fit the SED of the host galaxy. Similar to JADES\_954, this source is only slightly above the local $M_{BH}-M_{*}$ relation, with $M_{BH}/M_{*}=0.007$. From the BlackTHUNDER high-resolution spectrum, we derived $E(B-V)=0.12$, consistent with the value computed in \cite{Harikane23}. From the broken powerlaw fit to the PRISM spectrum, we got $\beta_{UV}=-1.97$ and $\beta_{opt}=-2.36$, therefore this source appears as a blue Type I AGN, even if  $\beta_{UV}$ is bluer than the typical UV slope of SDSS Type I QSO \citep{VandenBerk01}, possibly suggesting that the rest-UV is instead dominated by star-formation \citep{Saxena24}. It is also clearly extended morphologically, and the NIRCam RGB image revealed two close-by red and blue spots (not distinguishable in the {\it JWST} IFU cube due to the larger NIRSpec/IFU PSF). The effective radius of the Sersic-like component is estimated by \texttt{pysersic} to be $\sim 650$ pc. For this source, {\it JWST}/MIRI observations are also available in the F770W, F1000W, F1500W, and F2100W filters, and the source is detected in F770W and F1000W.

\subsection{NOEMA data} \label{sec:target_NOEMA}

The targets were observed with the IRAM interferometer NOEMA in intermediate B or BC configuration during the Winter 2024/25 semester (Project ID: W24EK, PI: H. Übler) using the PolyFiX correlator with 12 antennas in band 3. The \CIIfir fluxes were estimated starting from the targets' SFRs and assuming the $4<z<6$ \CII-SFR relation reported by \citep{Schaerer20}, derived from sources of the ALPINE survey \citep{LeFevre20, Faisst20}. The SFR of each source was estimated by assuming that the narrow component of the \Ha emission was due only to the host galaxy SF processes, by correcting it for dust attenuation and finally using the $L_{H\alpha}-$SFR relation derived from \cite{Kennicutt12}. The exposure times of the observations are reported in Table~\ref{tab:observations}. The excellent weather quality during the observations allowed us to achieve sensitivities even better than requested. JADES\_954 was observed on 27/28 December 2024 using MWC349 as a flux calibrator. JADES\_1146115 was observed on 3 January 2025 using LKHA101 as calibrator.  CEERS\_00397 was observed over three tracks on 2 January, 13 January and 16 March 2025 using as calibrator 2010+723 for the first track and MWC349 for the other two. All calibrations and the creation of uv table were performed using the \texttt{CLIC} package from the \texttt{IRAM GILDAS} software\footnote{\url{https://www.iram.fr/IRAMFR/GILDAS/}}. The resulting data cubes have native velocity resolution of $\sim 2.5$\kms and synthesized beam sizes with the following major$\times$minor axes: $0.82^{\prime\prime}\times0.75^{\prime\prime}$ (JADES\_954), $0.89^{\prime\prime}\times0.62^{\prime\prime}$ (JADES\_1146115), $0.75^{\prime\prime}\times0.65^{\prime\prime}$ (CEERS\_00397). All beam sizes are larger than the optical emission of these sources in the {\it JWST} images.

\begin{table}
\caption{Summary of the results of the NOEMA observations and of the host galaxy physical properties inferred from the \CII and dust continuum non detections (see Sect.~\ref{sec:target_NOEMA},~\ref{sec:Mgas_Mdust_ul})}
\centering
\setlength{\tabcolsep}{2pt}
\begin{tabular}{p{1.6cm} c c c c}

   & JADES\_954 & JADES\_1146115 & CEERS\_397 & Stack \\
   \hline \\

   Time & 8.7 & 5.2 & 6.4 & 20.3\\
   {[hrs]} & & & & \vspace{0.15cm}\\

   $\sigma_{[CII]}$ & 0.017 & 0.019 & 0.044 & 0.013\\
   {[\jykms]} & & & & \vspace{0.15cm}\\

   $\log L_{[CII]}$ & $<7.77$ & $<7.79$ & $<8.10$ & $<7.65$\\
   {[\Lsun]} & & & & \vspace{0.15cm}\\

   $\sigma_{cont}$ & 16 & 17 & 25 & 0.13\\
   {[$\mu$Jy]} & & & & \vspace{0.15cm}\\

   $\log L_{158\mu m}$ & $<43.80$ & $<43.82$ & $<43.91$ & $<43.65$\\
   {[erg $\rm s^{-1}$]} & & & & \vspace{0.15cm}\\
    \\
    \hline
    \\
    $\log(M_{\rm gas}/$\Msun)& $<9.26$ & $<9.29$ & $<9.59$ & $<9.13$\\
   {from \CII} & & & & \vspace{0.15cm}\\
   $\log(M_{\rm dust}/$\Msun)& $< 6.13$ & $<5.68$ & $<6.55$ & $< 6.01$\\
   {from \CII } & & & & \vspace{0.15cm}\\
   $\log(M_{\rm dust}/$\Msun)& $< 6.47$ & $<6.47$ & $<6.67$ & $< 6.34$\\
   {from cont.} & & & & \vspace{0.15cm}\\
   \\
    \hline
    
    \end{tabular}
    \tablefoot{{\it Top:} On-source NOEMA observing time, sensitivity and luminosity upper limits for both the \CII\ emission line and the underlying continuum, as described in Sect.~\ref{sec:target_NOEMA}. {\it Bottom:} Upper limits on the host galaxy gas and dust masses, as derived in Sect.~\ref{sec:CIIund}.}
    \label{tab:observations}
\end{table}

\section{Results and discussion} \label{sec:results}

\subsection{\CII and continuum non-detection} \label{sec:undetections}
The \CII emission is not detected at the 3$\sigma$ level in all three sources. In Appendix~\ref{sec:app_tentativeCII}, we report tentative \CII detections for JADES\_954 and JADES\_1146115, and we also discuss in detail why we did not consider them as reliable enough, or at least not as directly associated to the AGN. 

To derive the \CII 3$\sigma$ luminosity upper limit we proceeded as follows. We extracted 1D spectra from 100 random positions in the native cube using beam apertures, summing the flux of the pixels within the beam and dividing by the number of pixels per beam. We measured a median noise of the 1D spectra in the native velocity channels (2.5 \kms) of 1 mJy beam$^{-1}$, 0.91 mJy beam$^{-1}$, 1.93 mJy beam$^{-1}$, for JADES\_954 and JADES\_1146115, and CEERS\_00397, respectively. Then we rescaled the noise values to those measured re-binning the spectra at the velocity resolution corresponding to the expected FWHM of the \CII, assuming the same FWHM of the narrow optical emission lines. Then, we computed the \CII 3$\sigma$ flux upper limits assuming a box line shape \citep[with a width equal to the narrow-line FWHM, following][]{Walter09} and we derived the luminosity upper limits using Eq.~1 in \cite{decarli23}. The \CII $3\sigma$ flux and luminosity upper limits are reported in Table~\ref{tab:observations}.

\begin{figure}[h!]
\centering
	\includegraphics[width=0.9\columnwidth]{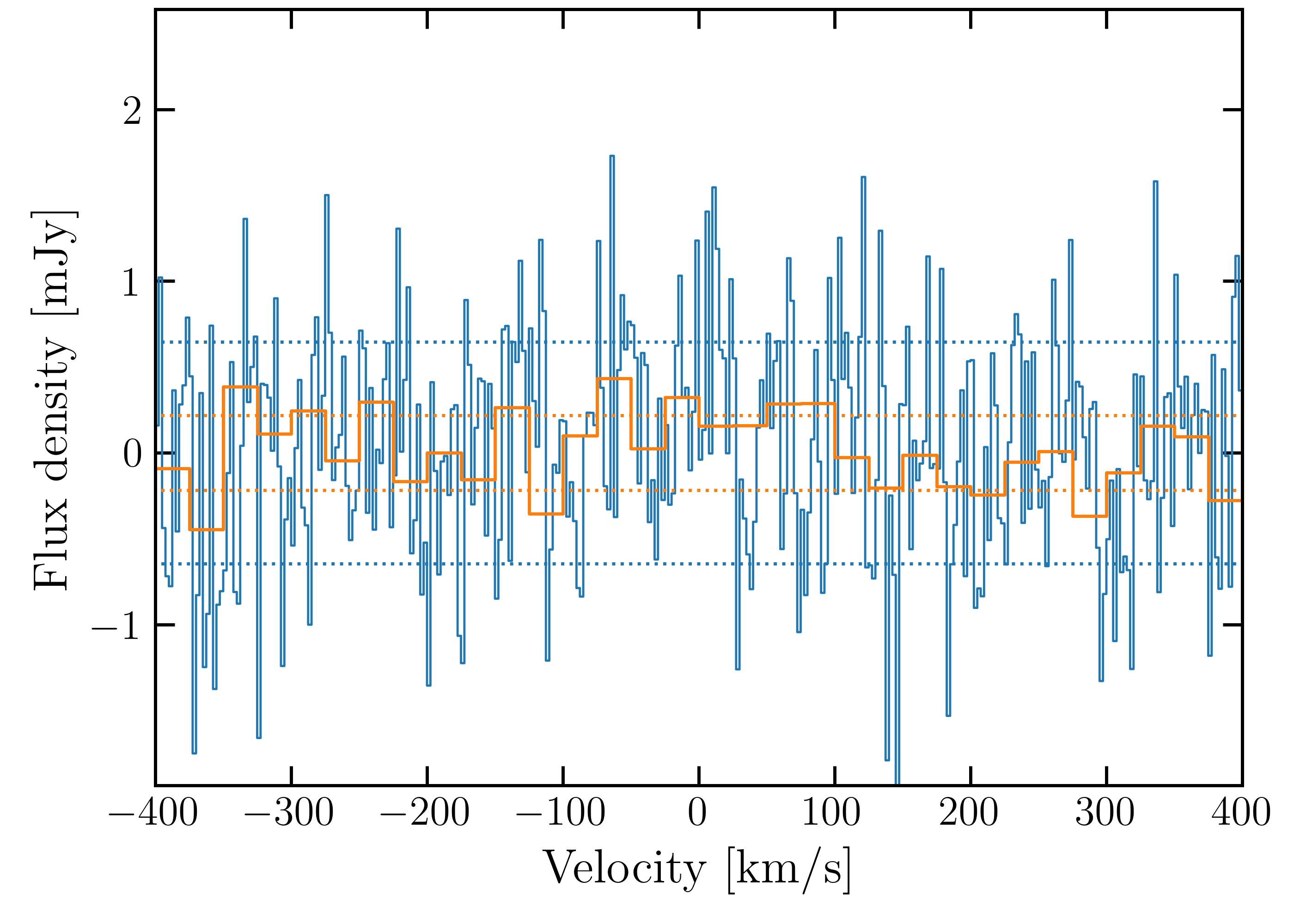}
    \includegraphics[width=0.7\columnwidth]{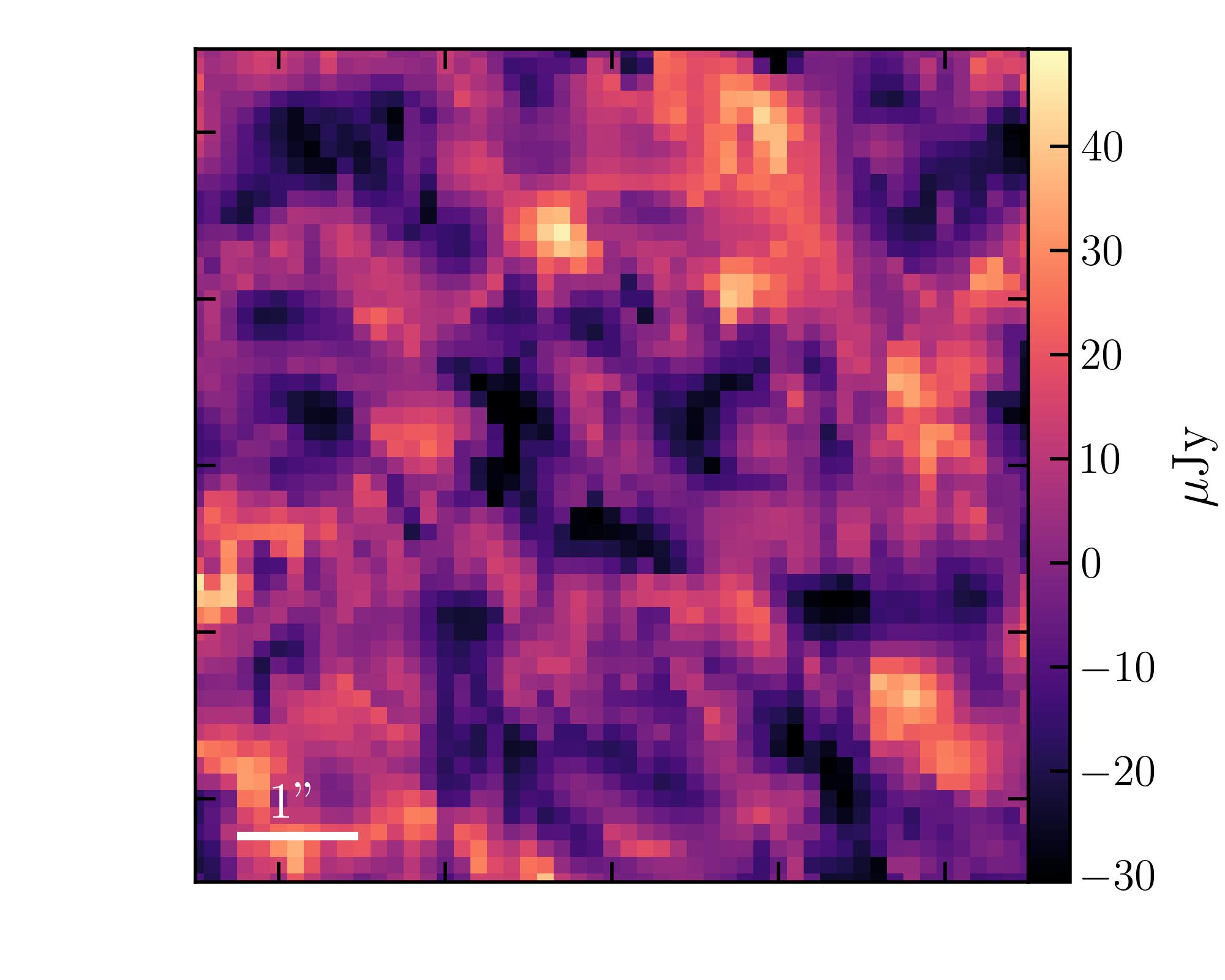}
    \caption{{\it Top:} median stack of the 1D spectra extracted at the position of the sources with beam-size apertures. The blue spectrum and the two blue horizontal dashed lines indicate the median stacked spectrum and rms at the native resolution of the cubes (2.5 \kms), while the orange lines refer to velocity channels of 25 \kms. {\it Bottom:} median stack map (50 pixels $\times$ 50 pixels) of the continuum maps of the three targets, revealing no detection in the center (the expected position of the targets).} 
    \label{fig:CII_stack}
\end{figure}

Given the  \CII non-detections, we performed a median stack of the 1D spectra extracted with beam apertures at the expected position of the sources and centered at the zero velocity corresponding to \CII. From the stacked spectrum, we measured an rms of 0.65 mJy beam$^{-1}$ for the 2.5 \kms\ binning. Taking the median FWHM of the three sources and the median redshift ($z=6.677$) we derived a $\log (L_{[CII]}/$\Lsun$)<7.65$. The stacked spectrum is shown in Fig.~\ref{fig:CII_stack}.\\

By collapsing both the upper side band (USB) and the lower side band (LSB) of the NOEMA observations, we also derived continuum maps, which did not reveal any detections down to a 3$\sigma$ level. To derive the 3$\sigma$ limits, we computed the rms from the median absolute deviation of the 2D map of each source. The continuum upper limits are reported in Table~\ref{tab:observations}. We also performed a median stack of the three continuum maps, following the same procedure outlined in \cite{Mazzolari26_radioJWST}. The stack did not reveal any detection, as shown in Fig.~\ref{fig:CII_stack}.

\subsection{Implications of \CII non-detection} \label{sec:CIIund}
\begin{figure}[h!]
	\includegraphics[width=1\columnwidth]{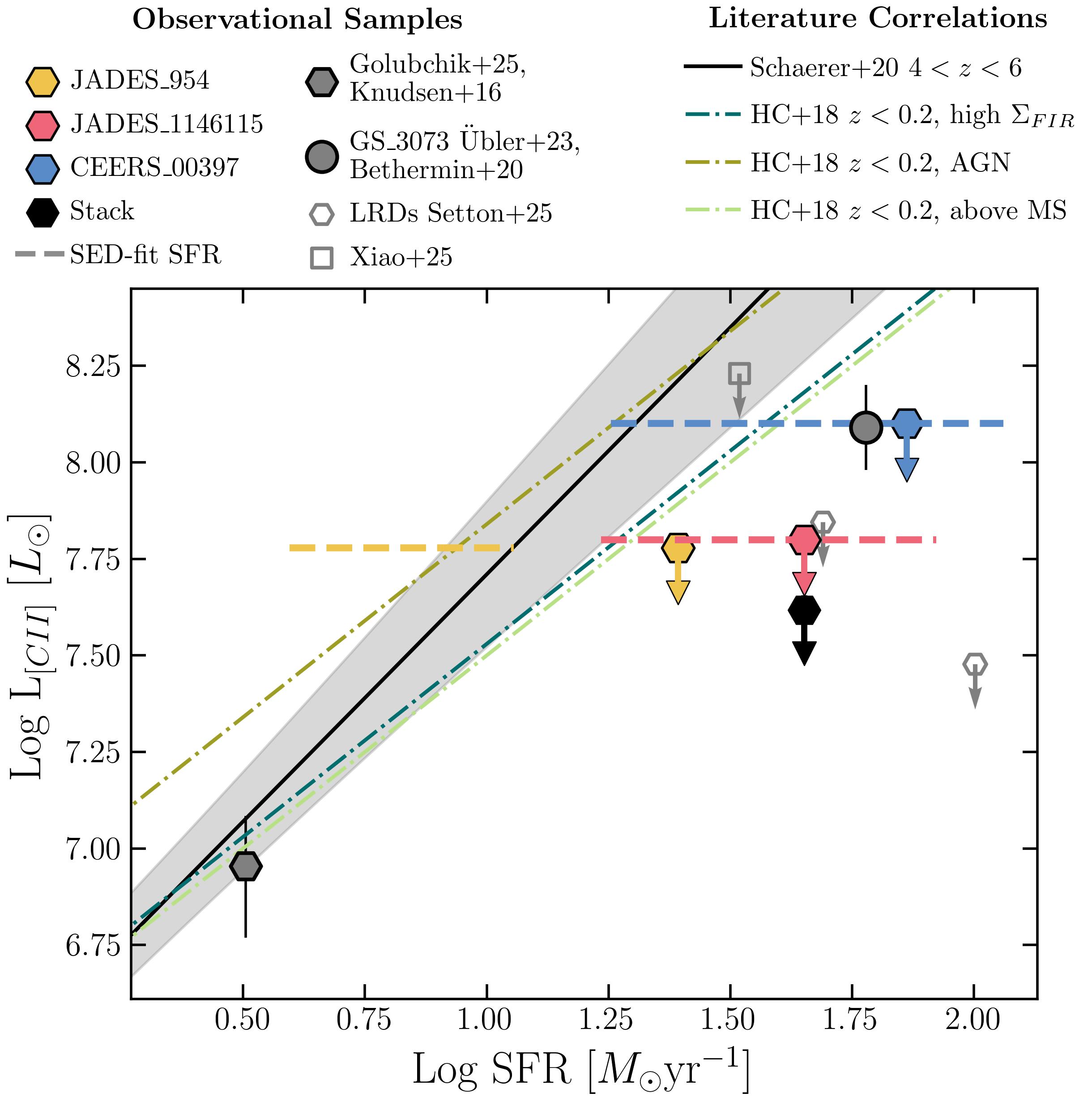}
    \caption{Position of our \CII $3\sigma$ upper limits (coloured hexagons and arrows) in the $L_{[CII]}-SFR$ plane. The black hexagon and arrow represent the results from the \CII stack. The black line and the gray shaded area represent the conservative relation from \cite{Schaerer20} and its scatter. The green dash-dotted lines represent correlations taken from \cite{HerreraCamus18} where the \CII is suppressed at a given SFR due to specific physical conditions, such as the presence of an AGN, galaxies in a starburst phase, or high radiation field intensities (i.e., high $\Sigma_{FIR}$). The horizontal dashed lines represent the range of SFR predicted by the SED-fitting decomposition presented in Sect.~\ref{sec:dustund}. In particular, the higher value corresponds to the instantaneous SFR and the lower one to the SFR averaged over the last 10 Myr. We also show upper limits of other LRD  and high-z {\it JWST} AGN undetected in \CII:  \cite[open hexagons:][]{Akins25_CI}, \cite[open square:][]{Xiao25_noema}. The gray hexagon shows a lensed LRD candidate  \citep{Golubchik25}, with a \CII detection reported by \cite{Knudsen16}. The gray circle shows GS\_3073 \citep{Vanzella06}, a blue compact and X-ray weak $z\sim 5.5$ AGN with bolometric luminosity comparable to those of our sources, with SFR derived from narrow H$\alpha$ by \cite{Ubler23} and \CII luminosity by \cite{Bethermin20}.
    } 
    \label{fig:CII_SFR}
\end{figure}
In Fig.~\ref{fig:CII_SFR} we show the upper limits derived from the \CII non-detections in the $L_{\CII}$-SFR plane. The upper limits on the \CII luminosity lie  $\sim 2-3\sigma$ below the $L_{\CII}$-SFR relation obtained by \cite{Schaerer20}. 
In Fig.~\ref{fig:CII_SFR} we also show the other spectroscopically confirmed {\it JWST} AGN with ALMA or NOEMA follow-up targeting \CII. Apart from the LRD photometric candidate A383-LRD1, where it was possible to detect faint \CII emission thanks to the large magnification of the source ($\mu\sim 16$), the other targets are similarly undetected in \CII, with luminosity upper limits close to those obtained in this work. While the \CII non-detection of all these sources can still be attributed to a scatter of the relation, there might be other reasons why these sources remain undetected. 

The SFRs of our targets were derived from the narrow \Ha emission, using the conversion from \cite{Kennicutt12} and assuming the line luminosity is dominated by SF from the host galaxy. Given that we are dealing with sources hosting Type I AGN, this might not be fully appropriate, given that the narrow \Ha component could also be powered by the AGN narrow-line region. However, the assumption appeared to be supported by the narrow FWHM of the sources $ 128<\rm FWHM/ \ km\ s^{-1}<240$, narrower than the typical narrow-line FWHM of Type I and Type II AGN \citep{mignoli19}. 
Additionally, the relation from \cite{Kennicutt12} does not account for the low metallicities of these sources. Assuming the more conservative relation by \cite{Theios19}, also used for high-z galaxies \citep{Shapley23, Pollock25}, would result in SFR estimates two times lower. The possibility that the SFRs derived from \Ha are overestimated is also partially supported by the range of SFR predicted by the new SED-fitting decomposition (presented in Sect.~\ref{sec:dustund}), which returned SFRs that are closer to the $L_{\CII}$-SFR relation of \cite{Schaerer20}. However, if this is the case, to reconcile the narrow \Ha luminosity with the SFR predicted by the \cite{Schaerer20} correlation at the given $L_{\CII}$ upper limits, the AGN contribution to the narrow \Ha must be $f^{AGN}_{NL \ H\alpha}>0.30, 0.59, 0.50$ for JADES\_954, JADES\_1146115 and CEERS\_397, respectively. While this might be reasonable for the blue Type I AGN (CEERS\_397), the LRD scenarios invoking dense and large covering factor gas distribution around the central BH do not expect the hard AGN ionizing radiation to contribute significantly to the narrow-line region, because most of it is absorbed on scales lower than the narrow-line region.

Given that our targets host AGN, another possible explanation for the \CII non-detection is that the AGN ionizing radiation could have impacted the formation of \CII. This could have happened as a combination of the destruction of small dust grains, reducing the photoelectric efficiency of the gas, and by converting a fraction of the $C^+$ ions to higher ionization states \citep{HerreraCamus18, Langer15}. This scenario has been particularly investigated by \cite{Langer15}, where, however, the main source of \CII suppression was the strong X-ray ionizing radiation. All the AGN investigated in this work are found to be observationally X-ray weak \citep{Maiolino24_X,Mazzolari25_CEERS} and not detected in the deep X-ray images of the Chandra Deep Field North or EGS field (having flux limits of $f_{0.5-keV}\sim 1-2\times 10^{-17} \cgs$ ). While it is still not clear whether this is due to an intrinsic weakness or to dense gas obscuration in the innermost part of the nuclear structure, in either case, it would be difficult for X-ray photons to interact strongly with the dust and gas of the host galaxies without producing any X-ray emission. Additionally, our three sources do not show strong high-ionization emission lines (such as \NeIV, \NeV, \HeII, \CIV), usually indicative of a hard AGN ionizing continuum, and that are also usually undetected in the general population of LRDs \citep{Inayoshi25_LRDreview,DEugenio25_irony,Torralba25} and {\it JWST} discovered AGN \citep{Zucchi26_highion, Lambrides24_Xweak,Maiolino2024_AGNsample}. The absence of these high-ionization emission lines has been interpreted as evidence for a screening gas structure covering the central engine and absorbing ionizing photons before they reach the narrow-line region or the host galaxy. It is worth noting that, to have $C^+$ ions (and therefore \CII emission), the required ionization potential is 11.26 eV (and therefore hydrogen ionizing photons are not needed), while an ionization potential of 24.38 eV is needed to convert $C^+$ into $C^{2+}$ (and therefore to deplete \CII emission by overionizing carbon species). In this context, it is interesting to note that GS\_3073, a compact, X-ray weak, and blue AGN at $z\sim5.5$ shows a compact but faint \CII detection (see Fig.~\ref{fig:CII_SFR}) even if its restframe UV and optical spectrum clearly shows several strong high-ionization emission lines \citep{Vanzella10, Grazian20, Barchiesi22, Ubler23}. This source has been recently presented as a prototypical `Little Blue Dot' (LBD) in \cite{Brazzini26_LBDLRD}, a population of high-z Type I AGN sharing with LRDs the X-ray weakness and compactness, but differing significantly in terms of continuum shape and presence of high-ionization emission lines. In conclusion, there is insufficient evidence in these sources to prove a general $C^{+}$ overionization, but this possibility cannot be fully ruled out.

We compared the sources' upper limits with the relations derived from samples observed as part of the SHINING survey and showing different physical properties, as outlined in \cite{HerreraCamus18}. Contrary to the relation of \cite{Schaerer20} (that considered a broad and uniform sample of star-forming galaxies (SFG) at $4<z<6$), these relations offer the possibility to investigate the impact of different physical conditions on the \CII production.  
\cite{HerreraCamus18} showed that while an AGN can significantly suppress \CII emission in the central few hundred parsecs of a galaxy by altering the heating and cooling balance of the interstellar medium, this effect becomes diluted on a global scale, as can be seen from the AGN relation in Fig.~\ref{fig:CII_SFR} lying very close to the SFG one by \cite{Schaerer20}. Instead, they found the main driver of \CII deficit to be the intense radiation field traced by $\Sigma_{FIR}$ or by the intense SFR, in particular in compact systems. Assuming the SFR from the \Ha line and using the SFR to far-infrared (FIR) luminosity conversion from \cite{Kennicutt12} we found $11.25<\log (\Sigma_{FIR}/$\Lsun $\rm pc^{-2} )<11.64$ by using the effective radii reported in Table~\ref{tab:sample}. These $\Sigma_{FIR}$ are already in the regime that allows thermal saturation of the \CII emission and therefore a deficit in the \CII vs SFR relation \citep{Lutz16,HerreraCamus18,Bisbas22_CII}. However, our observational results still lie well below the two relations derived in \cite{HerreraCamus18} for high $\Sigma_{FIR}$ and above main sequence (MS) sources.

\CII suppression in very dense systems may occour also as a consequence of its low critical density \citep[$\sim 10^3$ cm$^{-3}$][]{Carilli13}, that can go down to $\sim 50$ cm$^{-3}$ if the emission occurs in ionized gas regions, where the collision partners are electrons. Furthermore, in dense and gas-rich systems, the ability of hydrogen to shield itself from dissociation is increased, and the \CII emission is depleted \citep{Narayanan17}. 
As discussed in Sect.~\ref{sec:intro}, it is possible that these sources host heavily gas-enshrouded BHs, where the dense cocoon of gas can potentially reach very high densities ($n_H\sim10^9 cm^{-3}$), but large uncertainties remain on the density and properties of the gas in their host, given the millimeter non-detections.
Other scenarios for the lack of \CII include low-metallicity effects \citep[e.g.][]{Vallini25}, or the possibility that \CII is distributed on scales larger than those of the optical emission and therefore is characterized by a low surface brightness hard to detect even with deep mm observations \citep[a similar scenario was also proposed for the host galaxies of LRDs, see][]{Rinaldi25_LRD, Rinaldi25_Saguaro}.

The enhancement or depletion of \CII emission in AGN is a topic that has been widely debated in the literature \citep{Carniani20,Raouf25,HerreraCamus18, Langer15}. Indeed, there are also works reporting an excess of \CII emission in AGN host galaxies \citep{Smirnova-Pinchukova19} compared to what is expected for a standard SFG. In this case, the \CII enhancement was attributed to AGN outflows. However, the three sources analyzed in this work do not show the bright high velocity outflow signatures in ionized gas emission typical of luminous QSOs.

\subsection{Gas and dust mass upper limits from \CII and continuum non-detection} \label{sec:Mgas_Mdust_ul}
Works studying the evolution of the molecular gas fraction ($\mathrm{f_{molgas}=M_{molgas}/(M_*+M_{molgas})}$) consistently found a significant increase of $f_{\rm molgas}$ with redshift, with  $f_{\rm molgas}$ approaching unity already around cosmic noon \citep{Tacconi20, Genzel15, Freundlich19}.
From the \CII upper limits, we can derive an upper limit on the molecular gas mass using the correlation reported by \cite{Zanella18}, which provides a larger estimate of the gas mass compared to other relations accounting for a metallicity dependence of the $\alpha_{[CII]}$ \citep{Vallini25}. In this way we provide conservative upper limits on $M_{\rm gas}$. By taking $\alpha_{[CII]}=31\ \mathrm{M_{\odot}/L_{\odot}}$ from \cite{Zanella18}, we derived upper limits on the gas mass of $\log(M_{\rm gas}/M_\odot)<9.26$, $9.29$, $9.59$ for JADES\_954, JADES\_1146115 and CEERS\_397, respectively. From the stack \CII upper limit, we derived a median gas mass upper limit of $\log(M_{\rm gas}/M_\odot)<9.13$. With the metallicity-dependent conversion by \cite{Vallini25}, we got upper limits that are 0.3-0.5 dex lower. The molecular gas mass upper limit of JADES\_1146115 and CEERS\_397 are of the same order as the stellar mass estimated by \cite{Juodzbalis24} and \cite{Harikane23}, respectively. Instead, for JADES\_954 the molecular gas mass upper limit is quite low compared to the stellar mass derived by \cite{Maiolino2024_AGNsample} from the \texttt{BEAGLE} decomposition, being $\sim 1.5$dex lower than the stellar mass value reported there. Indeed, the stellar mass value reported by \cite{Maiolino2024_AGNsample} would imply $\mathrm{M_{gas}/M_*}<0.1$, implausible for the general population of galaxies $z\sim 6$ \citep[see][]{Genzel15,Tacconi20}, but also not expected for sources investigated in this work \citep[see][]{McClymont26_gasrichBH}. As we further discuss in Sec.~\ref{sec:dustund}, our new SED fitting decomposition returns a lower stellar mass for this object, reconciling this tension.

Assuming a dust-to-gas ratio (DGR) we can also derive an upper limit on the dust mass of the three sources. We assumed a linear evolution of DGR ratio with metallicty \citep{Draine07, RemyRuyer14, Sommovigo22}:
\begin{equation}
    \rm DGR=DGR_{\odot}\ \biggl(\frac{Z}{Z_{\odot}}\biggr),
\end{equation}
where DGR$_{\odot}$ is the Galactic value of 1/162, and $Z_{\odot}$ the solar metallicity. Taking the sources metallicities reported in Table~\ref{tab:sample}, we derived 3$\sigma$ upper limits on the dust mass of $\log(M_{dust}/$\Msun)$< 6.13, 5.68, 6.55$ for JADES\_954, JADES\_1146115 and CEERS\_397, respectively. From the gas mass upper limit obtained from the stack (and assuming the median metallicity of the sample) we derived $\log(M_{dust}/$\Msun)$< 6.01$.
We find the lowest dust mass for JADES\_1146115 which has also the lowest gas phase metallicity in our sample. This could potentially be related to recent accretion of low-metallicity gas.
Instead, using the metallicity-dependent DGR function obtained for $z \sim 0$ SFGs by \cite{Leroy11_gasconv}, we found $\sim$ two times larger dust masses. 
These individual values, thanks to the depth of the NOEMA observations, are consistent with those derived in the literature from a stack of similar objects at $z\sim 6$ \citep{Casey25_Mdust}.

We also derived the expected dust masses from the dust continuum upper limits. We fit the NOEMA non-detection by using a modified blackbody model. In particular, we used the following function:
\begin{equation}
S_{\nu, \mathrm{obs}} = M_{\mathrm{dust}} \,
\frac{1+z}{D_{\mathrm{L}}^{2}(z)} \,
\kappa_{\nu} \, \left[ B_{\nu}(T_{\mathrm{dust}}) - B_{\nu}\!\left(T_{\mathrm{CMB}}(z)\right) \right],
\end{equation}
where $S_{\nu,\mathrm{obs}}$ is the observed flux density at frequency $\nu_{obs}$, $M_{\mathrm{dust}}$ is the total dust mass, $\kappa_{\nu}$ is the dust opacity, \citep[for which we assumed the expression reported in][]{Carniani17,Costa26} and $D_{\mathrm{L}}(z)$ is the luminosity distance at redshift $z$. This equation gives the observed flux density measured against the CMB radiation field, represented by $B_{\nu}\!\left(T_{\mathrm{CMB}}(z)\right)$, where $T_{\mathrm{CMB}}(z) = 2.73\ (1+z)$ is the temperature of the CMB at redshift $z$.
To estimate the dust mass, we chose to fix the temperature at $T_{dust}=46$ K\footnote{We caution that some works found lower dust temperatures (specifically $T_{dust}\sim37$K) in one normal SFG at $z\sim 5.5$ \citep{Algera25_HZ10, Villanueva24}. However, it has also been demonstrated that AGN or QSO typically have $40~\rm K \lesssim T_{dust}\lesssim 70$~K \citep{Beelen06,Leipski13,Leipski14, Walter22}, and we therefore preferred to assume $T_{dust}= 46$K.} and $\beta=2$, taking the same cold dust temperature and $\beta$ of other studies of $z\sim6$ galaxies from the REBELS and ALPINE surveys \citep{Bowler24_IRX,Sommovigo22,Sommovigo20}. This returned upper limits of the dust masses of $\log(M_{dust}/$\Msun)$<6.47, 6.47, 6.67$, for JADES\_954, JADES\_1146115 and CEERS\_397, respectively. For the same parameters, we also inferred the upper limits on the far-infrared (FIR) luminosities of these sources to be: $\rm L_{FIR}<2.1\times10^{11}$ \Lsun, $<2.05\times10^{11}$ \Lsun, $<3.1\times10^{11}$ \Lsun. The upper limits to the dust masses obtained from the continuum non-detection are only 0.1 dex and 0.4 dex larger than the upper limits estimated above from the \CII upper limits for CEERS\_397 and JADES\_954, respectively, while it is 0.8 dex larger for JADES\_1146115. Using the contnuum stacked map, and considering the median redshift of the three sources, we derived an upper limit to the median dust mass of the sources $\log(M_{dust}/$\Msun)$< 6.34$, only 0.3 dex larger than what estimated from the stack \CII non detection.

\begin{figure}
	\includegraphics[width=0.95\columnwidth]{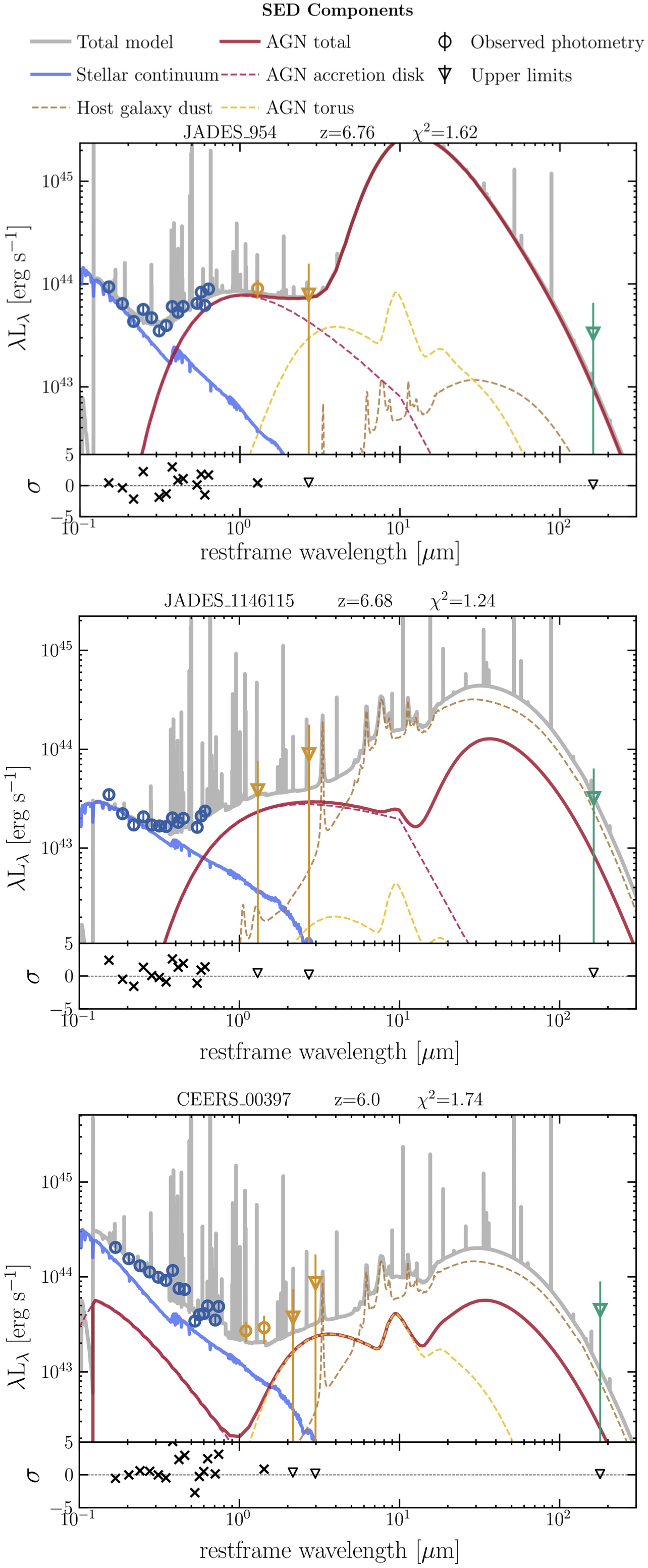}
    \caption{SED fitting decomposition of JADES\_954 (top), JADES\_1146115 (middle), and CEERS\_397 (bottom). {\it JWST}/PRISM data points are reported in blue, {\it JWST}/MIRI data in orange, and our new NOEMA upper limits in green. The positions of the upper limits (triangles) are reported at $2.5\sigma$, allowing the errors to go from zero to $5\sigma$. The blue line represents the stellar contribution, while the red solid line represents the AGN contribution (sum of obscured disk emission, torus emission, and polar dust emission). The red dashed line represents the contribution coming from the accretion disk emission, that for JADES\_954, and JADES\_1146115 is obscured in the polar direction. The brown line represents the host galaxy dust emission. The final model is reported in gray.}
    \label{fig:SED}
\end{figure}

\begin{table*}[h!]\label{tab:CIGALE_output}
\centering
\caption{Best-fit output parameters obtained from the \texttt{CIGALE} run.}
\label{tab:SED_output}

\begin{tabular}{lccc}
\hline
\textbf{Parameter} & \textbf{JADES\_954} & \textbf{JADES\_1146115} & \textbf{CEERS-00397} \\
\hline
\vspace{-0.2cm} 

\\

$\log(M_\star/M_\odot)$
& $8.28 \pm 0.33$
& $8.67 \pm 0.28$
& $8.75 \pm 0.24$ \\

$\log(M_{\rm gas}/M_\odot)$
& $8.03 \pm 0.76$
& $8.23 \pm 0.50$
& $8.26 \pm 0.41$ \\

$\log(M_{\rm dust}/M_\odot)$
& $7.36 \pm 1.08$
& $7.31 \pm 0.5$
& $7.53 \pm 0.64$ \\

instantaneous SFR [$M_\odot\,{\rm yr}^{-1}$]
& $11 \pm 10$
& $83 \pm 67$
& $114 \pm 56$ \\

SFR [10 Myr] [$M_\odot\,{\rm yr}^{-1}$]
& $4 \pm 0.7$
& $17 \pm 3$
& $18 \pm 2$ \\

Age main stellar population [Myrs]
& $192 \pm 168$
& $184 \pm 177$
& $255 \pm 168$ \\

\vspace{-0.2cm} 
\\

\hline
\vspace{-0.2cm} 

\\

$E(B-V)_{\rm AGN}$
& $0.96 \pm 0.13$
& $0.91 \pm 0.43$
& $0.14 \pm 0.04$ \\

Torus opening angle $\theta$ [deg]
& $10 \pm 4$
& $26 \pm 15$
& $32 \pm 12$ \\

Viewing angle $i$ [deg]
& $23 \pm 19$
& $22 \pm 25$
& $26 \pm 24$ \\

AGN fraction ($f_{\rm AGN}$)
& $0.75 \pm 0.05$
& $0.43 \pm 0.16$
& $0.22 \pm 0.05$ \\

AGN bolometric luminosity [erg\,s$^{-1}$]
& $(5.2 \pm 2.0)\times10^{45}$
& $(2.8 \pm 2.3)\times10^{44}$
& $(2.9 \pm 1.7)\times10^{44}$ \\

\hline
\\
New $M_{BH}/M_*$
& 0.28$_{-0.2}^{+0.4}$
&0.72$_{-0.4}^{+1}$
&0.03$_{-0.02}^{+0.05}$\\
\\
\hline
\end{tabular}
\tablefoot{We list host galaxy properties in the upper part and AGN properties in the lower part. The last line reports new estimates for $M_{BH}/M_*$ obtained using the new stellar mass estimates together with BH mass estimates from locally calibrated scaling relations. Uncertainties correspond to $1\sigma$ confidence intervals. }
\end{table*}

\subsection{Implications of the dust continuum non-detection} \label{sec:dustund}

We performed an SED-fitting decomposition to combine the rest-frame optical and UV light from {\it JWST} data with the NOEMA continuum non-detection and MIRI data. 
We used the {\it JWST}/NIRSpec integrated PRISM spectra available from the BlackTHUNDER and JADES surveys (see Fig.~\ref{fig:targets}) to extract the rest-frame optical and UV photometry by sampling the spectra extracted from apertures that include all the sources' light. In particular, we sampled the {\it JWST}/NIRSPec PRISM spectra (covering the observed wavelength range $0.65 \mu m<\lambda<5.6 \mu m$) in bins of 2500 \AA, which provides a good representation of the continuum shape without being too sensitive to noise fluctuations. We checked that the extracted fluxes are consistent with the NIRCam measurements using available NIRCam photometry. The SED-fitting decomposition was performed using \texttt{CIGALE}, and a detailed description of the modules and of the parameters explored is reported in Appendix \ref{sec:app_SEDfit}. In Table~\ref{tab:CIGALE_output} we also report the main output parameters, while the AGN and host galaxy emission decomposition are presented in Fig.~\ref{fig:SED}. 

For the two LRDs (JADES\_954 and JADES\_1146115) the macroscopic decomposition is similar: the rest-frame UV is dominated by stellar emission, while the AGN emission is suppressed in the rest-frame UV and starts to dominate at MIR wavelengths ($\lambda_{rest}\geq1 \mu m$). Instead for the blue AGN CEERS\_397, the code predicts an SED dominated by stellar emission and with a non negligible ($\sim 25\%$ in the $0.1 \mu m<\lambda_{rest}<5 \mu m$) contribution coming from a blue, almost unobscured AGN (as expected from the properties reported in Sect.~\ref{sec:targets_JWST}). A more detailed analysis of the AGN emission of these sources is discussed in Sect.~\ref{sec:AGN_fit}.

\texttt{CIGALE} assumes energy balance, and therefore the energy absorbed in the rest-frame optical and UV is then re-emitted in the FIR following the prescription of \cite{Draine14,Draine07}. This module accounts for diffuse dust emission heated by the general stellar population (cold dust) and also for dust tightly linked to star-forming regions (hotter dust). In the fit, we fixed the extinction of the optical emission lines, $E(B-V)$ to the values obtained from the narrow Balmer decrement in the {\it JWST} spectra of the sources and the stellar attenuation to be $0.44<E(B-V)_{star}/ E(B-V)_{lines}<1$. 

The instantaneous SFRs predicted by the code are compatible (within 1$\sigma$) with the values obtained from the narrow \Ha emission line (except for JADES\_954), and are mainly driven by SF burst predicted to occur in all the sources in the last 2-5 Myr. Instead, the SFRs averaged over the last 10~Myrs of star formation histories are $\sim 0.5-1$ dex lower, as shown by the dashed lines in Fig.~\ref{fig:CII_SFR} -- a potential explanation for the [CII] non-detections, as discussed in Section~\ref{sec:CIIund}. 

The stellar masses estimated from the SED fitting are lower than previous estimates in the literature, particularly for JADES\_954 and CEERS\_397. For JADES\_954, our new SED fitting predicts a stellar mass of $\rm \log(M_{*}/M_\odot)=8.3$, $\sim2$ dex below the value previously reported by \cite{Maiolino2024_AGNsample}, and just $\sim3$ times larger than the BH mass reported in Tab.~\ref{tab:sample}. Instead, for CEERS\_397, the stellar mass is $\rm \log(M_{*}/M_\odot)=8.7$, a factor of five lower than the previous estimate. Contrarily, for JADES\_1146115, the SED fitting predicts a stellar mass very similar to the one reported by \cite{Juodzbalis24}, where a careful AGN and host galaxy 2D decomposition was performed. These new stellar mass estimates alleviate the tension reported in Sect.~\ref{sec:Mgas_Mdust_ul} on the gas-to-stellar mass ratio when the literature stellar masses were considered.
 
These results demonstrate the importance of combining multi-wavelength constraints for a reliable derivation of global physical properties (see also in Sect~\ref{sec:MIRI-NOEMA}). At the same time, the SED-fitting decomposition we obtained including the MIRI data and the new NOEMA constraints, revealed (or confirmed for the case of JADES\_1146115) the overmassive nature of these sources and even increases the $M_{BH}/M_{*}$ ratios (estimated from locally-calibrated scaling relations), particularly for JADES\_954 that now shows $\mathrm{M_{BH}/M_{*}}\sim 0.3$. The new $\mathrm{M_{BH}/M_{*}}$ are reported in Table~\ref{tab:CIGALE_output}. However, we emphasize the uncertainty of this ratio, given the stellar masses uncertainties derived by \texttt{CIGALE} are $\sim 0.3-0.4$ dex and the systematic uncertainties of the relation used to derive $M_{BH}$ are $\sim 0.3-0.5$ dex \citep{Reines13,Reines15}.

The gas masses predicted by \texttt{CIGALE} are $\log(M_{\rm gas}/M_\odot)=8.03$, 8.32, 8.28 (respectively for JADES\_954, JADES\_1146115, and CEERS\_397), lower than the upper limits obtained in Sect.~\ref{sec:Mgas_Mdust_ul} from the \CII non-detection, and come with an uncertainty of $\sim$ 0.4-0.7 dex. The dust masses obtained from the code are usually more uncertain and are also slightly larger than the upper limits obtained in Sect.~\ref{sec:Mgas_Mdust_ul} from the dust continuum.

\subsection{AGN emission}\label{sec:AGN_fit}
The deep, multi-wavelength constraints from the {\it JWST}/PRISM spectrum, MIRI images, and NOEMA upper limits allowed us to provide a more complete picture of the AGN emission in these objects. From SED fitting, the AGN emission is predicted to be face on (as for a Type I AGN, with inclination relative to the line of sight of $16^{\circ} \lesssim i \lesssim 26^{\circ}$), consistent with the detection of broad emission lines. For the blue AGN (CEERS\_397), the code predicts the AGN contribution to come from an unobscured Type I AGN component, which, however, is not dominant over the stellar contribution across the whole SED. This was already suggested by the UV slope, larger than typically observed in Type I AGN, and closer to the intrinsic $\beta_{UV}$ of high-z SFG (see Sect.~\ref{sec:targets_JWST}). 
For the two LRDs, instead, the Type I AGN emission is expected to be obscured in the polar direction, with $E(B-V)\sim1$. Perhaps surprisingly, this result is consistent with the broad Balmer decrement observed in JADES\_954 and with the non-detection of the broad \Hb in JADES\_1146115. 
The light absorbed by the AGN polar dust (whose temperature was allowed to vary between $100$K and 300 K) is then re-emitted in the FIR, where the polar dust AGN emission dominates (or is comparable to) the host galaxy dust emission. The two LRDs are in GOODS-N, and this part of their SED could, in principle, be tested using Herschel constraints. However, these data are not deep enough to put constraints in the $10<\lambda_{rest}/\mu m<100$ range, but could in principle be explored by the future PRIMA facility \citep{Moullet25_PRIMAbook, McKinney_PRIMAlrd}. We highlight that polar dust obscuration in Type I AGN has already been reported not only for single objects but also for statistical samples at lower redshifts \citep{Lusso12, Elvis12_SEDty1cosmos, Stalevski16, Lyu18_polardust}. For example, \cite{Bongiorno12} suggested that the fraction of extincted sources (having E(B-V)$> 0.1$) among Type I AGNs is $\sim 40\%$. Therefore, the scenario we propose is not atypical and was also originally suggested to fit the continuum emission of other LRDs \citep{Wang25_LRDz3}.

An additional element that strongly supports the SED-fitting decomposition presented in this work is that, for all three sources, the AGN bolometric luminosities predicted by the SED-fitting are consistent within 1$\sigma$ with the values obtained in the literature from the scaling relations using the broad \Ha emission line. This is in contrast to the results by \cite{Greene25} for similar objects, where the authors stressed the importance of new bolometric corrections for LRDs and {\it JWST}-discovered AGN.    

It is worth noting that JADES\_1146115 and JADES\_954 (the two LRDs) show a particularly flat MIR SED. This property was already observed in other works and attributed to the lack of a significant warm dust component coming from the dusty torus \citep{Setton25_dust, Akins25_LRD, Casey24_LRDdust}. Also from our SED-fitting decomposition the dusty torus component is found to be subdominant at $1 \mu m<\lambda_{rest}<5 \mu m$ compared to the obscured accretion disk emission. In particular, the opening angle of the torus is expected to be small (i.e. $\theta \lesssim 26°$) and therefore only poorly illuminated by the central accretion disk. 

Recently, different AGN-related SEDs have been proposed to explain the red optical emission observed in high-z AGN discovered by {\it JWST}, particularly in LRDs.  For example, the gas-enshrouded AGN SED presented by \cite{Inayoshi2024_densegas} and observationally tested in \cite{Ji25_QSO1a, DEugenio25_irony}, or the BH-star modelling presented by \cite{Naidu25_BHstar, deGraaff25_BB, Kido25_BHstar}. The main prediction of the first model is the presence of absorption features in the broad Balmer lines up to the Balmer break, whose strengths depend primarily on the hydrogen volume and column densities. The main predictions of the BH-star model are instead the presence of a cool (at $T\sim 5000$K) partially ionized gas envelope that dominates the emission in the rest-optical and near-infrared, emitting similarly to a modified blackbody \citep{Kido25_BHstar,Liu25_BHstar}. While for CEERS\_397 these models are not consistent with its SED, given the blue optical and UV slopes, for JADES\_954 and JADES\_1146115 they might provide a plausible fit solution. Although both sources exhibit a turnover in their spectra at the Balmer limit, neither shows a strong one. Additionally, JADES\_954 shows a clear narrow \Ha absorption in its high-resolution spectrum, while no absorption features have been detected in the medium resolution spectrum of JADES\_1146115 (but it is still possible that absorptions are present but not resolved at $R\sim 1000$).
As for the BH-star model, we noticed that the obscured Type I AGN disk emission predicted by \texttt{CIGALE} resembles that of a blackbody and peaks at $\lambda_{peak,rest}\sim 0.7-2 \mu m$, almost the same range where most of the LRDs presented in \cite{deGraaff25_BB} are found to peak.
However, for the two LRD investigated in this work, there is no reason to prefer such a scenario over the more `classic' view presented and discussed in this section. The SED decomposition and the interpretation of the AGN emission presented in this work are instead consistent with what was found by \cite{Nikopoulos25}. There, analysing the Balmer decrements of multiple Balmer transitions in a sample of seven Type I AGN discovered by {\it JWST}, the authors found indications of a two-component scenario, where the broad lines (and possibly the optical/NIR emission) originate from a high column density AGN, while the narrow lines (and possibly the UV continuum) arise from a low-dust narrow line region or star-formation. A dust-driven interpretation of the properties of LRDs was also recently presented in \cite{Madau26_LRDLBD}.

In conclusion, the fit presented in this section provides a reliable and `classic' decomposition of the AGN and host galaxy emission of the two LRDs, without the need to invoke exotic scenarios. In particular, the LRD V-shape SED (and flat in the rest-NIR) can be explained, at least for these sources, as dominated by stars in the rest-UV, and by an accretion disk obscured in the polar direction in the rest-optical and rest-NIR, without a significant contribution from a dusty torus. This does not mean that the dusty torus is absent, but simply that it is confined to low opening angles, is not significantly illuminated, and therefore does not dominate the NIR SED with its reprocessed light.

\subsection{Importance of MIRI and NOEMA data}\label{sec:MIRI-NOEMA}
Adding the MIRI and NOEMA constraints to the SED fitting significantly improved the reliability of the results and the overall description of the sources compared to the sole rest-UV and optical information. We demonstrate this by performing a fit using the same parameter grid as for our fiducial fit, but without including the NOEMA upper limits or the MIRI data.

Excluding the MIRI constraints leads the code to add, in all three sources, a hot-dust component from the dusty torus, which dominates at $\lambda_{rest}>1\mu m$. As we instead showed in Sect.~\ref{sec:AGN_fit}, from the MIRI data and upper limits, the torus is not expected to dominate the emission at $1\lesssim\lambda_{rest}/\mu m\lesssim5$, because most of the light from the accretion disk does not intercept the torus opening angle. Additionally, without the MIRI data, the AGN component of the blue AGN CEERS\_397 is not fit as a blue Type I AGN, but rather as an obscured Type I AGN, similar to the other two LRDs.

Performing the fit excluding the NOEMA upper limits, we found for all sources the predicted 1.2 mm emission exceeding the limit set by our observations by more than a factor of 10, and in the case of CEERS\_397 even by a factor of $\sim 40$. This excess is mainly driven by a larger predicted AGN and galaxy host dust contribution in the mm bands. For the AGN, this is a consequence of a larger optical obscuration and a much larger intrinsic AGN luminosity, not consistent with the values reported in Table~\ref{tab:sample}. Instead, the dust masses computed without the NOEMA upper limits are a factor of 2, 7, and 3 larger for JADES\_954, JADES\_1146115, and CEERS\_00397, respectively. Given that the dust luminosity is fixed by the known attenuation, without the NOEMA limits, the code predicts a larger host galaxy cold dust component in these sources, which is instead ruled out by our continuum non-detections.

These results highlight the importance of multi-wavelength coverage to constrain physical properties through SED fitting. This is particularly relevant for the ongoing debate about the nature of distant and faint AGN, and for our understanding of the physical properties of their host galaxies.

\begin{figure}
	\includegraphics[width=1\columnwidth]{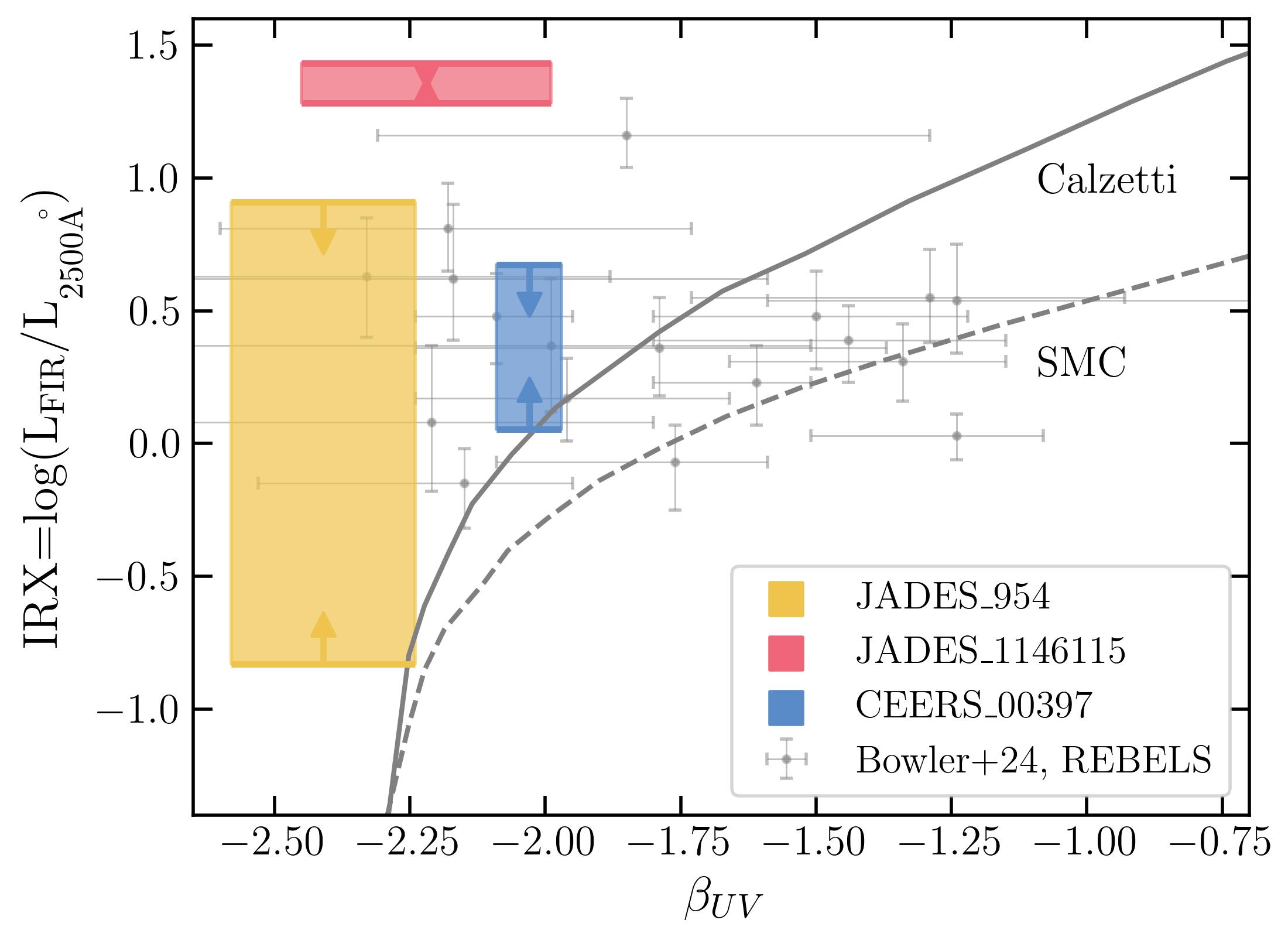}
    \caption{$IRX$ versus $\beta_{UV}$ of our targets. The upper and lower limits of the three shaded areas are given by the FIR luminosities from the FIR modified blackbody fit and the SED-fitting (energy balance), respectively. The width of the shaded area in the horizontal direction is given by the uncertainties on $\beta_{UV}$. The expected relation for Calzetti-like dust is shown by the grey solid line, while the expected relation from an SMC extinction curve is shown by the dashed line \citep{Meurer99, Reddy18a}. These models assume an intrinsic $\beta_0$ for the stellar population of $-2.3$, consistent with recent {\it JWST} results at $z>6$ \citep{Saxena24}. Grey points show $6<z<8$ detections from the REBELS survey by \cite{Bowler24_IRX}.}
    \label{fig:IRX}
\end{figure}
\subsection{Indications of complex dust-UV geometry} \label{sec:IRX}
In the analysis presented in Sect.~\ref{sec:CIIund} and in Sect.~\ref{sec:dustund} we computed the FIR emission of our targets in two different ways: the FIR luminosity estimate provided by \texttt{CIGALE} and the upper limit provided by the modified blackbody fit at fixed $T_{dust}=46$~K. Comparing these two FIR luminosities can provide valuable insight into the dust properties of these sources and, potentially, on how the dust is distributed relative to the optical light. The FIR luminosity computed by \texttt{CIGALE} assumes energy balance and considers both the contribution from attenuated starlight and from the attenuated nebular emission (nebular lines plus nebular continuum). This result can be considered as a lower limit to the total FIR luminosity of a source, given that the energy absorbed by dust in the rest-UV and optical has to be re-emitted in the FIR, and this amount is determined by the obscuration of the narrow lines measured from the narrow Balmer decrement. However, more complex geometries between the UV-emitting regions and dust might actually determine larger FIR emission than predicted by the energy balance scenario. On the other hand, the modified blackbody fit provides the maximum possible cold dust luminosity given the observed upper limit at the chosen dust temperature, and therefore it can be considered as an upper limit to the FIR luminosity.
With these constraints, it is useful to place the three sources in the IRX versus $\beta_{UV}$ plane \citep{Meurer99}, where IRX=$\log(L_{FIR}/L_{1500\AA})$, while $\beta_{UV}$ represents the rest-frame UV powerlaw slope. The location of the source in this plane provides an indication of the dust properties, the attenuation or extinction law characterizing the sources, and the geometry of the FIR-emitting regions. We measured $\beta_{UV}$ from the available {\it JWST} PRISM spectra in the wavelength range $1250<\lambda/\AA<2500$ by fitting the spectrum with a simple power-law model after masking regions of 100 \AA\ around strong emission lines. 

In Fig.~\ref{fig:IRX} we show the position of the three sources in the IRX-$\beta_{UV}$ plane.
The lower and upper limits obtained for IRX with the two values of FIR luminosity identify a relatively large region for both JADES\_954 and CEERS\_397, whereas the two values are close to each other for JADES\_1146115. The latter is also the source with the largest attenuation, as determined from the narrow Balmer decrement. In all three cases, the region spanned by the upper and lower limit of the IRX is above the track corresponding to the \cite{Calzetti00} attenuation law, and is located in a region where usually sources show a displacement between the UV and FIR emitting regions \citep[see][]{Bowler24_IRX,Villanueva24}. More precisely, according to the predictions by \cite{Popping17}, this is a region of the IRX-$\beta$ plane where the UV emission might be partially screened due to the presence of "holes" in the dust screening medium. In particular, JADES\_954 and CEERS\_397 are compatible with dust-free emission fractions ranging from 20\% to 100\%. Instead, the position of JADES\_1146115 is compatible with lower non-obscured fractions, i.e., <5\%-10\%. This result suggests that the upper limits derived from the NOEMA observations indicate a complex geometry or a patchy distribution of the dust relative to the rest-UV emission, which is rather unexpected given the small sizes of the sources (see Sect.~\ref{sec:targets_JWST}). While the current non-detections do not allow any further investigation, we caution that deviations in the IRX-$\beta_{UV}$ diagram towards the bluer, upper part of the parameter space have also been differently interpreted. In particular, \cite{Popping17} showed that an increased level of turbulence in the screen of dust can lead to lower values for both $\beta$ and IRX, such that galaxies move towards locations in the IRX–$\beta$ plane above the relation for a uniform dust screen (so that for a fixed IRX, galaxies become bluer). A similar IRX parameter space can also be reproduced by stars mixed in between the screen of dust (rather than placed in front of it), even though in this case the $\beta_{UV}$ slopes are expected to be redder \citep[see ][]{Goldader02, Nordon13}. In particular, the extreme position of JADES\_1146115 is hard to interpret without accounting for a patchy dust distribution.

\section{Conclusion} \label{sec:conclusion}
In this work, we performed an in-depth analysis of the implications of the mm non-detections of three {\it JWST}-discovered Type I AGN at $6<z<7$ observed for $\sim 5-10$~h with NOEMA targeting \CIIfir. JADES\_1146115 and JADES\_954 are LRDs, while CEERS\_00397 is a blue Type I AGN. In the analysis, we leveraged multi-wavelength information from: the {\it JWST}/PRISM spectra, which provide optimal characterization of the rest-frame UV and optical continuum; the MIRI images, which provide useful constraints in the rest-frame NIR and MIR; and the NOEMA data, which constrain the FIR SED and the gas properties. The main results are summarized as follows:
\begin{itemize}
    \item No \CII emission is detected in our targets, also after stacking their spectra. We estimated a 3$\sigma$ upper limit on the \CII emission, which locates them $\sim 2\sigma$ below the most conservative \CII-SFR relation derived from \cite{Schaerer20}. These non-detections might be due to an overestimation of the sources' SFRs, to AGN ionizing radiation, or to density or metallicity effects. In the first scenario to reconcile the \CII upper limits with the scaling relations a contribution of the AGN to the narrow \Ha line $>30-50\%$ is needed, also for the two LRDs.\\
    
    \item From the \CII non-detections, we derived conservative upper limits of the host galaxy gas masses of $M_{gas}\lesssim 2-4\times10^9$\Msun\ and of their dust masses of $M_{dust}\lesssim 5-50\times 10^5$ \Msun, consistent with what was found by other works performing stacking analysis. Using a modified blackbody model to fit the continuum non-detections, we instead found slightly larger upper limits on the dust mass $M_{dust}\lesssim 3-5\times10^6$ \Msun.\\
    
    \item The three sources are also undetected in the 158$\mu$m continuum, also after stacking their maps. We performed a detailed SED-fitting analysis using NOEMA, MIRI, and NIRSpec data and derived revised physical properties for the three sources. We find host galaxy stellar mass 2 dex lower than reported in the literature for JADES\_954 and 0.7 dex lower for CEERS\_00397, while consistent with previous results for JADES\_1146115. Using our new stellar mass estimates together with BH mass estimates from locally calibrated scaling relations, we found $M_{BH}/M_{*}=0.28, 0.72, 0.03$ for JADES\_954, JADES\_1146115, and CEERS\_00397, respectively. Despite the large uncertainties affecting these estimates, these findings suggest that our targets host overmassive BHs. \\
    
    \item The SED fitting decomposition predicts the AGN emission of CEERS\_397 to be consistent with that of a blue Type I AGN, but subdominant compared to stellar emission. Instead, for the two LRDs, their peculiar SED shapes can be explained by star formation in the rest-UV and by a Type I AGN accretion disk obscured in the polar direction, in the rest-optical and NIR. This `classical' AGN configuration can explain the flat optical-NIR slope (observed in many LRDs), and its expected emission is very similar to the one predicted by the dense gas envelope characterizing the BH-star model. Our SED decompositions returned AGN bolometric luminosities consistent with those predicted by scaling relations based on \Ha\ luminosity. Our analysis suggests that the interpretation of the observed properties of high$-z$ AGN discovered by {\it JWST} (including LRDs) is diverse, and different physical models may need to be considered to explain the data.\\
    
    \item 
    We placed our targets in the IRX-$\beta$ plane, finding that they occupy a region were sources might be characterized by complex and patchy geometries between the dust and the UV emitting region.
    
\end{itemize}
The upper limits on \CII obtained through our NOEMA observations provided
important constraints on the physical properties of our targets, both on the gas and dust content of their host galaxies and on their AGN emission. 
Yet deeper observations, or larger stack experiments, may be required to detect such distant, compact AGN with $mm$ observations to finally provide a clear and unbiased view of their host galaxies.


\begin{acknowledgements}
We thank Gan Luo for the help in the reduction of the NOEMA data. We acknowledge useful conversations with Christina Eilers, Gene Leung, Roberto Gilli, Marcella Brusa and Marco Mignoli. GM and H\"U acknowledge funding by the European Union (ERC APEX, 101164796).
H\"U thanks the Max Planck Society for support through the Lise Meitner Excellence Program.
NMFS, GT, JC, CB, JME, CP, LL acknowledge funding by the European Union (ERC, GALPHYS, 101055023).
AJB acknowledges funding from the “FirstGalaxies” Advanced Grant from the European Research Council (ERC) under the European Union’s Horizon 2020 research and innovation program (Grant agreement No. 789056).MP acknowledges support through the grants PID2021-127718NB-I00, PID2024-159902NA-I00, and RYC2023-044853-I, funded by the Spain Ministry of Science and Innovation/State Agency of Research MCIN/AEI/10.13039/501100011033 and El Fondo Social Europeo Plus FSE+.
EB acknowledges funding from INAF “Ricerca Fondamentale 2024” (GO grant ``A JWST/MIRI MIRACLE: Mid-IR Activity of Circumnuclear Line Emission’' and mini-grant 1.05.24.07.01). TN acknowledges support from the Deutsche Forschungsgemeinschaft (DFG, German Research Foundation) under Germany’s Excellence Strategy - EXC-2094 - 390783311 from the DFG Cluster of Excellence "ORIGINS”.
Views and opinions expressed are those of the authors only and do not necessarily reflect those of the European Union or the European Research Council Executive Agency. Neither the European Union nor the granting authority can be held responsible for them.

\end{acknowledgements}

\bibliographystyle{aa}
\bibliography{literature}

\begin{appendix} 

\section{Tentative line detections in JADES\_954 and JADES\_1146115}\label{sec:app_tentativeCII}

\begin{figure}
	\includegraphics[width=1\columnwidth]{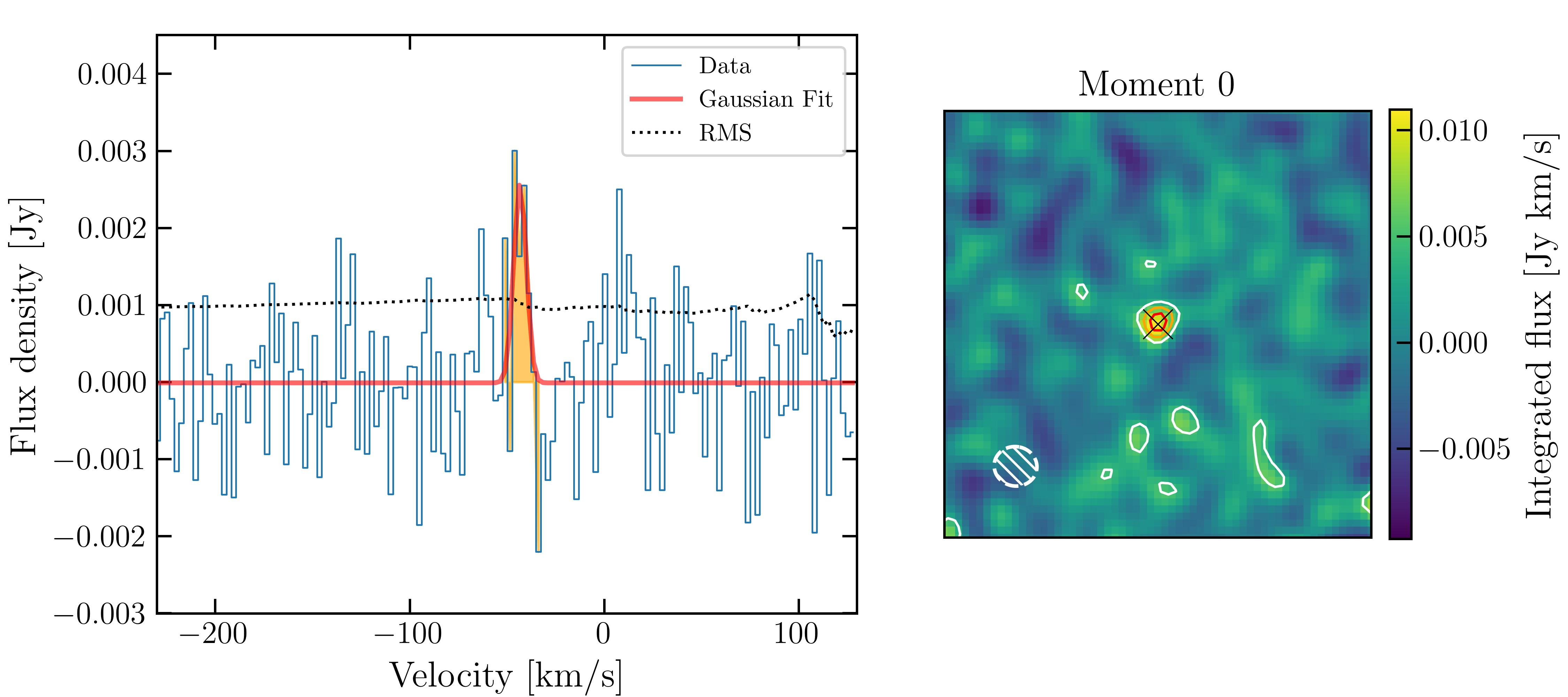}\\
    \includegraphics[width=1\columnwidth]{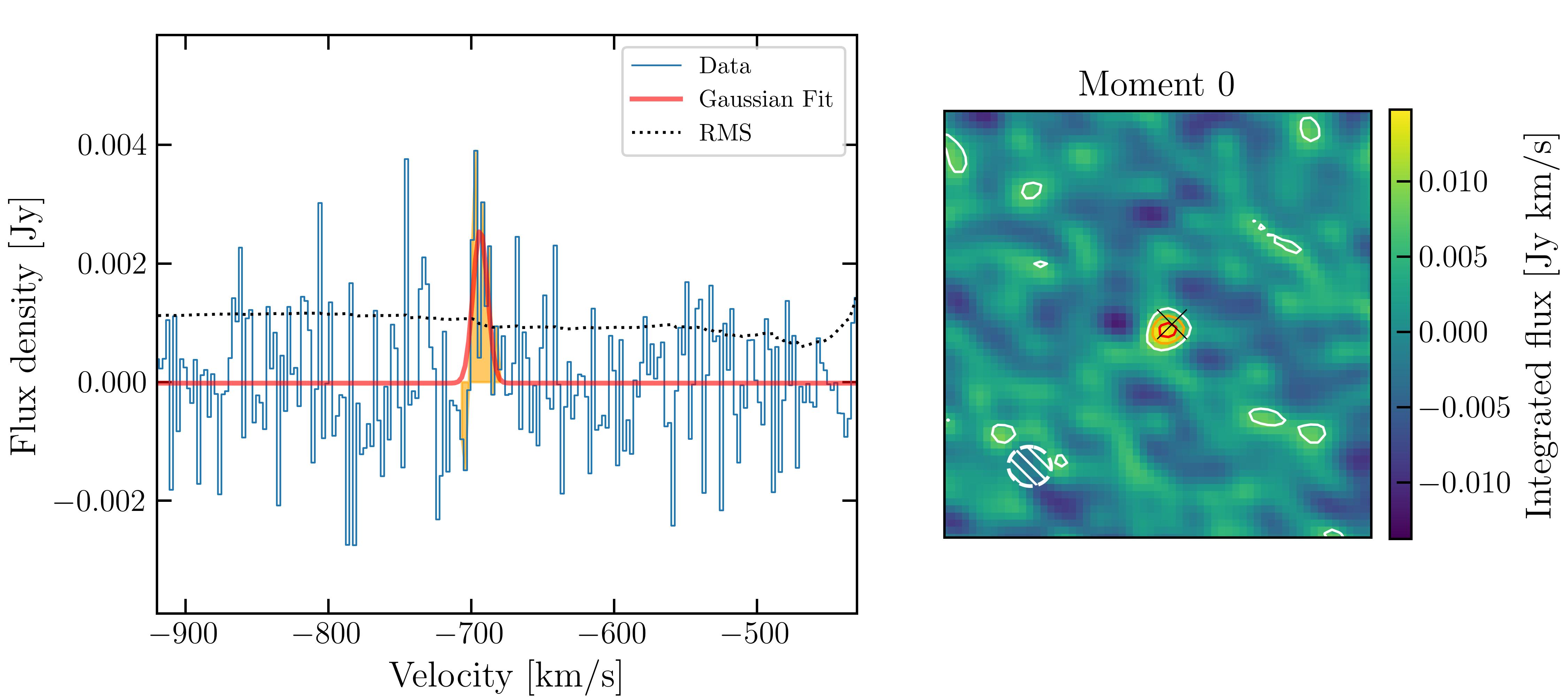}\\
    \includegraphics[width=1\columnwidth]{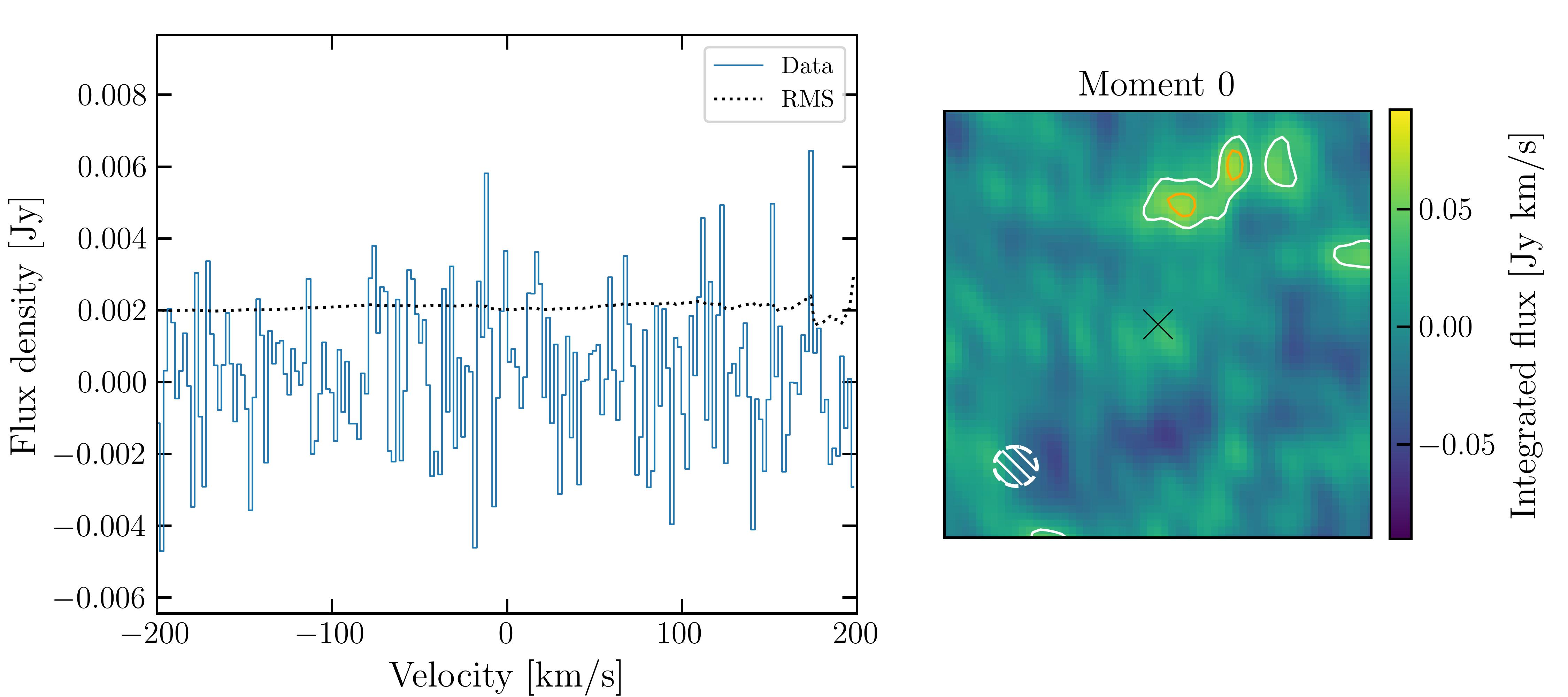}
    \caption{Tentative detections of \CIIfir for JADES\_954 (upper panel) and JADES\_1146115 (middle panel), and the non- detection for CEERS\_00397 (lower panel). On the left we show the 1D spectra extracted from a beam aperture at the position of the source and the Gaussian fits of the tentative lines. On the right, we show the moment-0 map at the velocity of the tentative lines (for JADES\_954 and JADES\_1146115), collapsing channels within $\pm 1\sigma$, where $\sigma$ is the width returned by the Gaussian fit. For CEERS\_00397 the moment-0 map was created collapsing the channels within $\pm 100$ \kms\ from the zero velocity. The white, orange, and red contours mark the 2$\sigma$, 3$\sigma$, and 4$\sigma$ noise level, respectively. The black cross marks the position of the source in the {\it JWST} optical images.}
    \label{fig:CIItentdet}
\end{figure}

As discussed in Sect.~\ref{sec:CIIund}, we could only derive upper limits on the \CIIfir emission at the location and redshift of our targets. However, we found tentative \CII detections near the systemic redshift in two sources: JADES\_954 and JADES\_1146115. The two tentative detections are shown in Fig.~\ref{fig:CIItentdet}.

None of these two tentative line detections appears to be significant enough when compared with the distribution of positive and negative peak flux densities obtained from the moment-zero maps, even though with integrated $S/N\sim 4.5$, and falling in the low-probability tail of the histogram of noise fluctuations. This result was confirmed by tests using the line-detection codes \texttt{interferopy} \citep{Walter16_interferopy} and \texttt{MF3D} \citep{Pavesi18_MF3D} with cubes at the native velocity resolution (2.5 \kms) and with a 10 times rebinning. Furthermore, even though the tentative lines are exactly at the position of the two sources, there are two additional indications suggesting that the lines may be spurious. First, they are both shifted in velocity compared to the nominal redshift of the source, -43 \kms\ for JADES\_954 and -693 \kms\ for JADES\_1146115. While for JADES\_954 this shift would still be consistent with the redshift of the source derived from high-resolution {\it JWST} spectroscopy within a couple of sigmas, for JADES\_1146115 the redshift uncertainty is not large enough to justify the velocity shift of \CII without invoking an actual shift of the cold gas emission relative to the ionized emission. Second, the FWHM of both lines are extremely narrow: 8 \kms\ for JADES\_954 and 12 \kms\ for JADES\_1146115. While a narrow FWHM might be expected given the low stellar masses of the two sources, these \CII FWHM are more than 1 dex lower than the FWHM of the ionized gas, which are 128 \kms\ for JADES\_954, and 180 \kms\ for JADES\_1146115. 

Assuming for a moment that these are real line detections, given the unresolved NOEMA observations (but still with a sub-arcsecond beam size), it is possible to obtain an upper limit on the dynamical mass of the host galaxies (using the same approach as in \cite{Ubler23} and taking as $R_e$ a mean between the half-major and half-minor axes of the beam). This computation gives $\log(M_{dyn}/M_\odot)\lesssim7.67$ for JADES\_954, and $\log(M_{dyn}/M_\odot)\lesssim8.12$ for JADES\_1146115, both lower than the stellar mass derived in Sect.~\ref{sec:dustund}, and two (three) times lower than the BH estimated from the \Ha line and reported in Table~\ref{tab:sample} for JADES\_954 (JADES\_1146115). This further indicates that these tentative line detections are in fact spurious, or at least not directly associated with the AGN.

However, for the case of JADES\_954 we note the small, blue-ish source extension to the North-East which is apparent in the NIRCam imaging (Fig.~\ref{fig:targets}) and in our size analysis (Sect.~\ref{sec:targets_JWST}). It is possible that the tentative \CII emission shown in the top panel of Fig.~\ref{fig:CIItentdet} and close to the systemic velocity of JADES\_954 is indeed real and associated to this bluer emitter, which could correspond to a low-mass satellite. Such faint, blue emission has been observed around several {\it JWST}-AGN \citep[e.g.][]{Matthee23, Ji2024a, Golubchik25,Rinaldi25_LRD,Baggen2026_bluecomp} and indeed in some cases could be associated to \CII\ detections (for instance in the cases of GS\_3073: \citealp{Bethermin20_ALPINE, Barchiesi22, Ubler23, Ji2024a}; ZS7: \citealp{Pentericci16, Uebler2024b}; A383-LRD1: \citealp{Knudsen16, Golubchik25}; or the $z\sim2$ source `Saguaro': \citealp{Freundlich19, Rinaldi25_Saguaro}; Mazzolari et al.~in prep.).

\section{\texttt{CIGALE} SED fitting}\label{sec:app_SEDfit}
The SED fitting was performed using a large parameter grid and accounting for AGN emission. The parameters used for the SED fitting are reported in Table~\ref{tab:ciagle_parameter}. 
We used delayed star-formation history (SFH) models, which can reproduce both early- and late-type galaxies, with an additional term that allows for a recent burst of star formation. We provide a range of ages of the main stellar population varying between the age of the universe at the redshift of the source and 10~Myr, and we allow the burst to happen within the last 10~Myr and to contribute up to 20\% of the total mass budget. We adopted stellar templates from \cite{bruzual03}, and a \cite{Chabrier03} initial mass function. We consider two different metallicities for the stellar populations (solar and 0.2 solar) and a separation between the young and old stellar populations between 10 and 500~Myr. We also include the nebular emission module with gas-phase metallicities of solar and 0.2 solar. This module is computed by \texttt{CIGALE} in a self-consistent manner using a grid of \texttt{Cloudy} photoionization calculations \citep{Ferland13}. For the attenuation of the stellar continuum emission, we use the \texttt{dustatt\_modified\_starburst} module with a Small Magellanic Cloud extinction law \citep{Pei92}. We select a narrow range of $E(B-V)_{\text{line}}$, the attenuation of the nebular emission lines, to match the result obtained from the spectral analysis (see Table~\ref{tab:sample}) and then set two possible ratios between $E(B-V)_{\text{line}}$ and $E(B-V)_{\text{star}}$. We also include dust emission in the infrared following the model by \cite{Draine14}, as reported in Sect.~\ref{sec:dustund}. 

For the AGN contribution, we employed the \texttt{skirtor2016} \citep{Stalevski16} module updated by \cite{Yang20}, which has been demonstrated to be reliable in studying various aspects of AGN \citep[e.g.][]{Mountrichas22, Lopez23, Yang23, Mazzolari25_CEERS,Mazzolari26_radioCTK} and include the possibility of Type I AGN obscuration via polar extinction, contrary to the \cite{fritz2006agnmodel} model. The SED produced by the AGN combines emission from the accretion disk, torus, and polar dust. The accretion disk, responsible for the UV-optical emission in the central region, is parametrized according to \cite{Schartmann05}, allowing for a deviation of the intrinsic slope in the $0.125<\lambda_{rest}/\mu m<10$ according to the parameter $\delta$ ($f_{\lambda}\propto\lambda^{-1.5\pm \delta}$). We allowed $-1<\delta<1$. 
Photons from the accretion disk can be obscured and scattered by dust in the vicinity, within the torus and/or in the polar direction. For the torus, the \texttt{skirtor2016} module employs a clumpy two-phase model \citep{Stalevski16}, based on the 3D radiative-transfer code SKIRT. For the polar dust component, we used the Small Magellanic Cloud (SMC) extinction curve, recommended for AGN observations \citep[e.g.][]{Bongiorno12}. For the polar extinction amplitude, we chose a range $0.05<E(B-V)<1.5$. The code also assumes energy conservation for the polar dust that absorbs the AGN emission, and we allowed the polar dust temperature to vary between $ 100$ and $300$ K.
We allowed the opening angle of the AGN torus to vary between 10°$<\theta<$40° and the AGN line of sight inclination 0°$<i<$90° (where values 0°$<i<$90°-$\theta$ correspond to Type I AGN, while 90°-$\theta<i<$ 90° to Type II AGN). Finally, we let the AGN fraction (defined as the ratio between the AGN luminosity and the total galaxy luminosity between 0.13 and 5$\mu$m) range from 0.1 (weak AGN contribution) to 0.9 (dominant AGN emission).

\begin{table*}[!h]\label{tab:ciagle_parameter}
   \caption{Main input parameters for SED fitting with {\sc CIGALE} (Section \ref{sec:dustund}).  \label{tab:sedfitting}}
   \centering
   \small
   \begin{tabular}{clcccccc}
   \hline\hline
   Module & Parameter & Symbol & Values \\
   \hline\xrowht[()]{5pt}
                                          & Stellar $e$-folding time & $\tau_{\rm{star}}$ [$10^6$ yr] & 10, 50, 100, 200,  500 \\
Star formation history                    & Stellar age      & $t_{\rm{star}}$ [$10^6$ yr]     & 10, 50, 100, 200, 300, 500, 700 \\
$[\rm{SFR} \propto t \exp(-t/\tau)]$      & Age of the burst                     & $t_{\rm{burst}}$ [$10^6$ yr]    & 1, 5  \\
                                          & Mass fraction of the burst population     & $f_{\rm burst}$                      & 0.0, 0.01, 0.1, 0.2 \\
\hline\xrowht[()]{5pt}

Single stellar population      & Initial mass function      & --                                              & \citet{Chabrier03} \\
$[$\cite{bruzual03}$]$         & Metallicity                     & $Z$          &            0.004, 0.02 \\
                              & separation between the young and the old star                      & sep\_age [$10^6$ yr]          &            10, 200, 500 \\

\hline\xrowht[()]{5pt}

\multirow{1}{*}{Nebular emission}  & Gas metallicity            & Z$_{\rm gas}$ &  0.004, 0.02 \\
                                  
\hline\xrowht[()]{5pt}
                                  
Dust attenuation                  & colour excess of the nebular lines    & $E(B-V)_{lines}$    & set to the values reported in Table~\ref{tab:sample} \\
$[$\cite{Pei92}$]$                & Fractional ISM attenuation    & $E(B-V)_{lines}/E(B-V)_{star}$   & 0.44, 1  \\

\hline\xrowht[()]{5pt}
                            & Mass fraction of PAH                                & q$_{\rm pah}$     &   0.47, 2.5, 3.9  \\
Galactic dust emission      & Minimum radiation field                             & U$_{\rm min}$     & 0.1, 1.0, 5, 10, 25, 40        \\
$[$\cite{Draine14}$]$             & Slope in $dM_{\rm{dust}} \propto U^{-\alpha}dU$     & $\alpha$          &  2.0 \\
                           & Illuminated fraction [Umin--Umax]              & $\gamma$          & 0.02, 0.1, 0.5  \\
\hline\xrowht[()]{5pt}

                           &  Edge-on optical depth at 9.7$\mu$m      & $\tau_{9.7}$     &      7        \\
                           &  Slope of radial dust density profile    &    --            &      1.0          \\
                           &  Slope of polar dust density profile     &    --            &      1.0          \\
                           &  Torus opening angle (from equator)      &   $\Theta$       &    0, 40            \\
AGN disk+torus+polar dust: &  Ratio outer/inner torus radius          &   --             &     20           \\
              &  Viewing angle (from vertical axis)      & $i$              &   0, 50, 90            \\
$[$\cite{Stalevski16}, \\
\cite{Yang20}$]$   &  AGN fraction                          &     $f_{\rm AGN}$      &    0.2, 0.3, 0.4, 0.5, 0.6, 0.7, 0.8, 0.9    \\
                           &  Extinction law for polar dust           &     --           &     SMC                      \\
                           &  $E(B-V)$ in the polar direction         &  --              &     0.03, 0.1, 0.2, 0.5, 1, 1.5         \\
                           &  Temperature of polar dust               &      [K]          &     100, 300                    \\
                           &  Emissivity of polar dust                &  --              &        1.6                  \\
\hline
   \end{tabular}

   \end{table*}

\end{appendix}

\end{document}